%% file: main.tex
\documentclass{natureEdit}
\usepackage{gensymb}
\usepackage{mathtools,bm}
\usepackage{xcolor}
\usepackage{graphics}
\usepackage{subcaption}
\usepackage[capitalise]{cleveref}
\usepackage[normalem]{ulem} %provides strike-through

\makeatletter
\let\saved@includegraphics\includegraphics
\AtBeginDocument{\let\includegraphics\saved@includegraphics}
\renewenvironment*{figure}{\@float{figure}}{\end@float}
\makeatother
\input{subfiles/notation}

\title{Seeing Around Corners with Edge-Resolved Transient Imaging}

\author{Joshua Rapp$^{1,2,\dagger}$, Charles Saunders$^{1,\dagger}$, Juli{\'a}n Tachella$^{3,\dagger}$, John Murray-Bruce$^{1,4}$, Yoann Altmann$^{3}$, Jean-Yves Tourneret$^{5}$, Stephen McLaughlin$^{3}$, Robin M. A. Dawson$^{2}$, Franco N. C. Wong$^{6}$ \& Vivek~K Goyal$^{1}$}

\begin{document}

\maketitle

\begin{affiliations}
 \item Department of Electrical and Computer Engineering, Boston University, 1 Silber Way, Boston, Massachusetts 02215, USA
 \item Charles Stark Draper Laboratory, Inc., 555 Technology Square, Cambridge, Massachusetts 02139, USA
 \item School of Engineering and Physical Sciences, Heriot-Watt University, Edinburgh, EH14 4AS, UK
 \item Department of Computer Science and Engineering, University of South Florida, 4202 E. Fowler Avenue, Tampa, Florida 33620, USA
 \item INP/ENSEEHIT-IRIT-TeSA, University of Toulouse, 31071 Toulouse Cedex 7, France
 \item Research Laboratory of Electronics, Massachusetts Institute of Technology, 77 Massachusetts Avenue, Cambridge, Massachusetts 02139, USA.
 \item[$^\dagger$] These authors contributed equally.
\end{affiliations}

\input{subfiles/00_abstract.tex}

\input{subfiles/01_intro.tex}

\section*{Results}
\subsection{Acquisition Methodology.}
\input{subfiles/02_acquisition.tex}

\subsection{Light Transport Model.}
\input{subfiles/03_lightTransport.tex}

\subsection{Reconstruction Approach.}
\input{subfiles/04_reconstruction.tex}

\subsection{Experimental Reconstructions.}
\input{subfiles/05_results.tex}

\section*{Discussion}
\input{subfiles/06_conclusions.tex}

\input{subfiles/07_methods.tex}

\section*{References}
\bibliography{subfiles/ref}
\bibliographystyle{naturemag}

\begin{addendum}
 \item We thank F. Durand, W. T. Freeman, J. H. Shapiro, A. Torralba, G. W. Wornell, and S. W. Seidel for discussions. 
 This work was supported
 in part by the US Defense Advanced Research Projects Agency (DARPA) REVEAL Program under contract number HR0011-16-C-0030,
 in part by a Draper Fellowship under Navy contract N00030-16-C-0014,
 in part by the US National Science Foundation under grant number 1815896,
 in part by the Royal Academy of Engineering under the Research Fellowship scheme RF201617/16/31,
 in part by the Engineering and Physical Sciences Research Council (EPSRC) (grants EP/T00097X/1 and EP/S000631/1), and
 in part by the MOD University Defence Research Collaboration (UDRC) in Signal Processing.
% \item[Author contributions] 
% J.R., C.S., J.M.-B. and V.K.G. conceived the acquisition method. 
% J.R. and C.S. derived the light transport model. 
% J.R. designed and performed the experiments. 
% C.S. derived the noise model and performed simulations. J.T. developed the reconstruction algorithm. 
% J.R., C.S., J.T., J.M.-B., and V.K.G. discussed the data. 
% Y.A., J.-Y.T., S.M., R.M.A.D., F.N.C.W., and V.K.G. supervised the research. 
% All authors contributed to the manuscript.
%\item[Competing Interests] The authors declare that they have no
%competing interests.
% \item[Correspondence] Correspondence and requests for materials
%should be addressed to \\
%V.K.G.~(email: v.goyal@ieee.org).
\end{addendum}

\end{document}

\maketitle

\etocsettocstyle{\subsection*{This supplemental document contains the following sections:}}{\noindent\rule{\linewidth}{.4pt}}
\localtableofcontents

% Transient light transport modeling
\input{SuppSubfiles/model_intro}
\input{SuppSubfiles/facetModel}
\input{SuppSubfiles/rotateFacet}

\input{SuppSubfiles/occludedFacet}
\input{SuppSubfiles/ceilingModel}

\input{SuppSubfiles/wedgeModel}

% Algorithm
\input{SuppSubfiles/skellapop}

% Observation model comparison
% \clearpage
\input{SuppSubfiles/comparison_observation_model}

% Experimental details
\input{SuppSubfiles/implementFigs}

% Raw data
\input{SuppSubfiles/data}

% Results
\input{SuppSubfiles/experiments}

% Robustness
\clearpage
\input{SuppSubfiles/robustness}

\bibliographystyle{unsrt}
\bibliography{SuppSubfiles/supplement}

%% file: subfiles/notation.tex
\newcommand{\Frac}[2]{{{#1}/{#2}}}  % an "inert" form of \frac

\def\Vecb{\mathbf{b}}

\def\Vech{\mathbf{h}}

\def\Vecm{\mathbf{m}}

\def\Vecu{\mathbf{u}}
\def\Vecv{\mathbf{v}}

\def\Vecy{\mathbf{y}}

\def\mv{\Vecv}
\def\mh{\Vech}

\def\fr{f_{\rm r}}
\def\nb{n_{\rm b}}
\def\nl{n_\ell}
\def\rl{r_\ell}
\def\ro{r_{\rm o}}
\def\rs{r_{\rm s}}
\def\tbin{t_{\rm bin}}
\def\tdw{t_{\rm dwell}}
\def\th{t_{\rm h}}
\def\tr{t_{\rm r}}

\def\ximin{\xi_{\rm min}}
\def\ximax{\xi_{\rm max}}

\def\rmid{r_{\rm mid}}
\def\Delt{\Delta_t}
\def\Dth{\Delta_\theta}

%% file: subfiles/00_abstract.tex
\begin{abstract}
Non-line-of-sight (NLOS) imaging is a rapidly growing field seeking to form images of objects outside the field of view, with potential
applications in search and rescue, reconnaissance, and even medical imaging.
The critical challenge of NLOS imaging is that diffuse reflections scatter light in all directions, resulting in weak signals and a loss of directional information.
To address this problem, we propose a method for seeing around corners that derives angular resolution from vertical edges and longitudinal resolution from the temporal response to a pulsed light source.
We introduce an acquisition strategy, scene response model, and reconstruction algorithm that enable the formation of 2.5-dimensional representations -- a plan view plus heights -- and a 180$\degree$ field of view (FOV) for large-scale scenes. 
Our experiments demonstrate accurate reconstructions of hidden rooms up to 3 meters in each dimension. 
\end{abstract}

%% file: subfiles/01_intro.tex
Significant strides have been made over the past decade to enable imaging beyond the line of sight%
\cite{Bouman2017,Baradad2018,Saunders2019,Seidel2019,Yedidia2019,Thrampoulidis2017,Xu2018,Kirmani2009,Velten2012,Heide2014,Buttafava2015,Gariepy2016,Pediredla2017,OToole2018,Ahn2019,Heide2019,Lindell2019a,Liu2019,Pediredla2019,Xin2019,Maeda2019,Lindell2019b}.
However, all of the proposed methods must contend with a pair of challenges resulting from diffuse reflections:  the scattering of light destroys directional information and diminishes the intensity as the inverse-square of the distance.
For line-of-sight (LOS) imaging (e.g., conventional photography or lidar), the effect of the diffuse reflection of the illumination toward a detector is minor because the directionality of light can be preserved through focused illumination or detection, and
the
radial falloff 
is often only a problem at large stand-off distances.
The difficulty of NLOS imaging arises because of the cascaded diffuse reflections\cite{Velten2012}.
For passive NLOS imaging methods\cite{Bouman2017,Baradad2018,Saunders2019,Seidel2019,Yedidia2019}, ambient illumination bounces off a hidden scene and a relay surface (usually a visible wall or floor) before reaching a detector.
Actively-illuminated approaches incur at least three diffuse reflections: from a visible relay surface to the hidden scene, off the scene, and back from the relay to the detector\cite{Kirmani2009,Velten2012,Heide2014,Buttafava2015,Gariepy2016,Thrampoulidis2017,Xu2018,Pediredla2017,OToole2018,Ahn2019,Heide2019,Lindell2019a,Liu2019,Pediredla2019,Xin2019}.
Some NLOS imaging approaches aim to avoid diffuse reflections entirely by using modalities (e.g., thermal\cite{Maeda2019}, acoustic\cite{Lindell2019b}, or radar\cite{Scheiner2019}) operating at long wavelengths, at which optically-rough surfaces appear smooth.
While the directionality and signal strength are preserved through one or more specular reflections, such methods measure physical properties other than the optical reflectance, the resolution is lower due to the longer wavelengths, and the specular reflections can lead to confusion in distinguishing between direct and indirect reflections.

The first challenge for optical NLOS imaging is how to recover directional information from 
diffusely reflected
light.
One
approach is to use active optical illumination and time-resolved sensing, for which 
the high-resolution transient information constrains the reconstruction of possible light directions to a more feasible inverse problem\cite{Velten2012,Heide2014,Buttafava2015,Gariepy2016,Pediredla2017,OToole2018,Ahn2019,Heide2019,Lindell2019a,Liu2019,Pediredla2019,Xin2019}.
Methods that do not capture temporal information\cite{Bouman2017,Baradad2018,Saunders2019,Seidel2019,Yedidia2019,Thrampoulidis2017,Xu2018,Torralba2014} -- especially those using passive illumination -- require a different approach to recovering directionality, such as by taking advantage of occlusions that block certain light paths and reduce directional uncertainty, similar to coded-aperture imaging or reference structure tomography\cite{Brady2004}.
Other methods use temporal differences between measurements to remove the contributions from static scene components, enabling improved resolution of an object in motion\cite{Bouman2017,Gariepy2016,Klein2016}.

The second main challenge is handling extremely low levels of informative light for macroscopic scenes.
Although active methods typically use single-photon detection and acquire transient information from many repeated illuminations, the hidden scene is still generally limited to around 1 meter from the relay surface.
Examples that form images of objects at the greatest distances or with the lowest acquisition times %simply
increase the laser illumination power to levels that are not eye-safe (e.g., 1~W average optical power at 532~nm)\cite{Lindell2019a,Liu2019}.
Passive, intensity-only methods generally have less radial fall-off but more uncertainty due to uninformative ambient light, so reconstructions are typically limited %in dimension
to 2D images of table-top scenes\cite{Saunders2019} or 1D traces showing objects' angular positions with respect to a vertical edge\cite{Bouman2017,Seidel2019}.

This work recovers large-scale 3D NLOS images with a large field-of-view (FOV) by combining  concepts from both active and passive methods for separating contributions mixed by diffuse reflection
into a novel acquisition configuration.
We call our method \emph{edge-resolved transient imaging}
(ERTI) because we take advantage of occlusions from vertical edges, in addition to using single-photon-sensitive, time-resolved acquisition.
Our method introduces a dual configuration to existing ``corner cameras'' by scanning a light source along an arc on the ground plane around a vertical edge, thereby controlling which portion of the hidden space is illuminated, and detecting light from a single spot.
Combining pulsed illumination and time-resolved, single-photon sensing yields measurements of the transient response of the illuminated scene.
Because of the novel acquisition geometry, differences between histograms of photon detection times
from adjacent illumination spots localize the transient response to that of a hemispherical wedge.
Light propagation modelling leads to closed-form expressions for the temporal response functions of planar facets extending from the ground plane for a given distance, height, orientation, and albedo.
Bayesian inference employing a tailored Markov chain Monte Carlo (MCMC) sampling approach reconstructs planar facets for each wedge and accounts for prior beliefs about the structure of likely scenes.

%% file: subfiles/02_acquisition.tex
Vertical edges, such as those in door frames or at the boundaries of buildings, are ubiquitous and have proved useful for passive NLOS imaging\cite{Bouman2017,Seidel2019}.
In so-called ``corner cameras'', a conventional camera images the ground plane 
where it intersects with the vertical edge of a wall separating visible and hidden scenes.
Whereas light from any visible part of the scene can reach the camera's FOV, the vertical edge occludes light from the hidden scene from reaching certain pixels, depending on their position relative to the vertical edge.
This work introduces two changes to how vertical edges enable NLOS imaging.
First, rather than using global illumination and spatially-resolved detection, we propose a Helmholtz reciprocal dual 
of the edge camera, in which the illumination is scanned along an arc centered at the edge, and 
a bucket detector aimed beyond the edge collects light from both the visible and hidden scenes.
Second, we use a pulsed laser and single-photon-sensitive, time-resolved detection instead of a conventional camera.

\begin{figure}
    \centering
    \includegraphics[width=\linewidth]{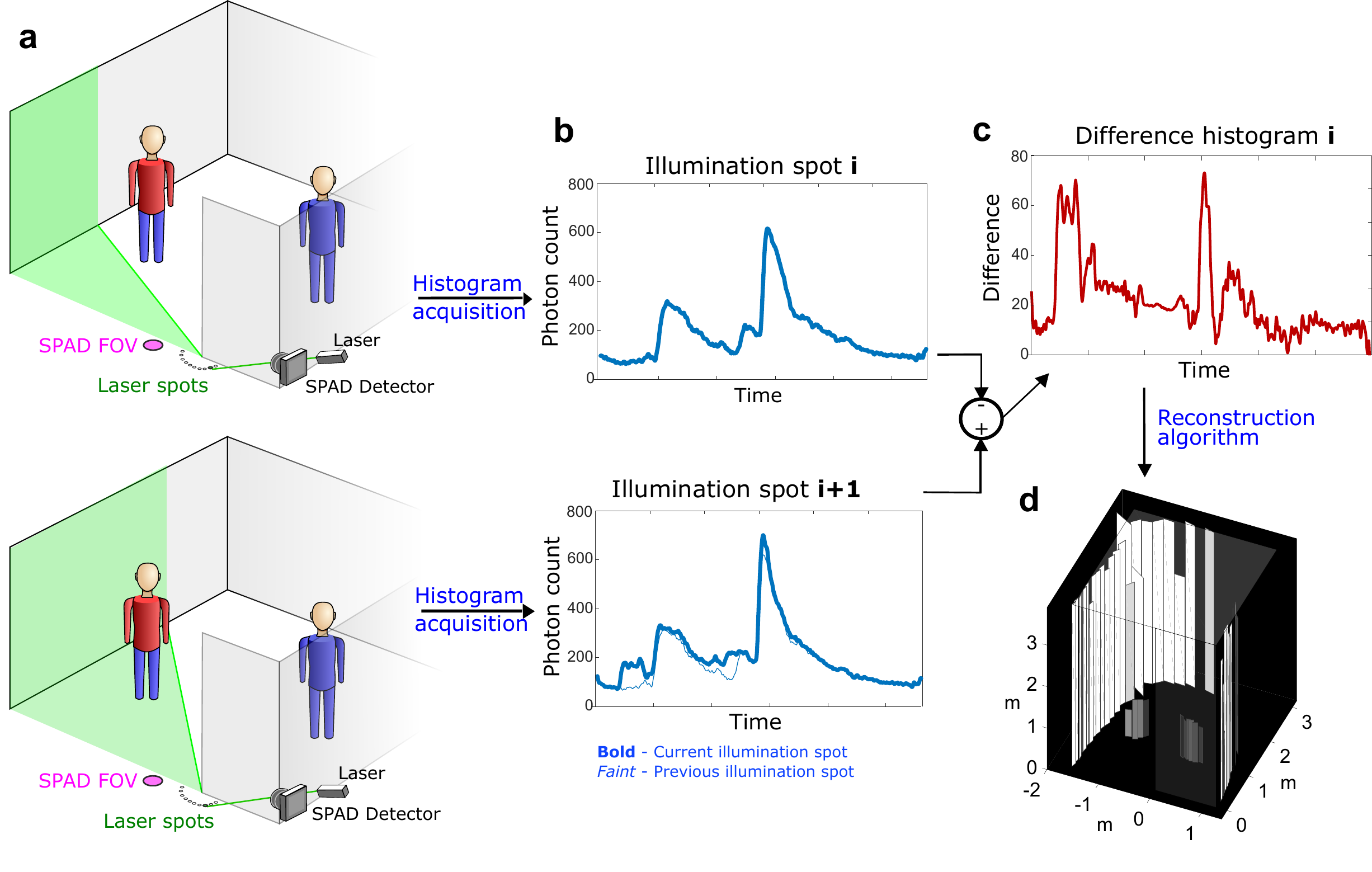}
    \caption{\textbf{ERTI scenario and procedure.} \textbf{a}, Positions along an arc centred on the occluding wall edge are illuminated sequentially by a pulsed laser.
    \textbf{b}, A histogram over time of photon detections 
    is collected for each laser illumination position. The measured histograms contain photons reflected from both the hidden scene and the visible side.
    \textbf{c}, Taking differences between sequential histograms 
    on average yields 
    returns originating only from a small wedge within the hidden scene.
    \textbf{d}, A hidden area reconstruction using the collection of histogram differences.
    }
    \label{fig:methodology}
\end{figure}

The ERTI acquisition methodology is illustrated in Fig.~\ref{fig:methodology}.
A 532-nm laser at 120~mW average optical power sequentially illuminates 45 spots evenly spaced in angle $\theta$ from $0$ to $\pi$ radians along a semicircle of radius 1.5~cm  
centred around the edge of the wall. 
The laser light illuminates
an increasing fraction of the hidden scene as the illumination spot moves along the arc toward the hidden area.
Each spot $i$ is repeatedly illuminated with picosecond-duration pulses at 20-MHz repetition rate for a preset dwell time.  
Light from each pulse bounces off the Lambertian ground plane and scatters in all directions,
reflecting from surfaces in
both visible and hidden portions of the scene.
A single-photon avalanche diode (SPAD) detector is focused at a small spot 
roughly 20~cm
beyond the vertical edge, enabling collection of light from the entire hidden scene for each illumination spot.
After each pulse, a time-correlated single photon counting (TCSPC) module connected to the SPAD records photon detection times with 16-ps resolution, forming a histogram $\Vecm_i$ of those photons reflected back to the SPAD\@. 
To prevent the direct reflection from overwhelming the much weaker light from the hidden scene, temporal gating is implemented to turn on 3~ns after the direct reflection from the ground reaches the SPAD.

The detected light intensity for spot $i$
includes contributions
$\mh_i$ from the hidden scene,
$\mv_i$ from the visible scene, and
$\Vecb$ from the background (a combination of ambient light and dark counts).
The background is assumed to have constant intensity over the duration of the acquisition, and because the illumination arc radius is small,
the visible scene contribution is 
approximately constant over all  spots,
i.e., $\mv_i \approx \mv_j$ for all $i, j$.
However, illuminating sequentially along an arc will change the parts of the hidden scene that are illuminated.
More 
precisely, a larger area 
of the hidden scene is illuminated as $i$ increases, so $\mh_{i+1} = \mh_i +\Vecu_i$, where  $\Vecu_i$ is the component of the histogram contributed by the portion of the scene illuminated 
from spot
$i+1$ but not 
from spot
$i$, and $\Vecu_0 = \mathbf{0}$ because only the visible scene is illuminated from the first laser spot.
The key idea behind ERTI is that this new contribution can be isolated -- thereby regaining NLOS directionality -- 
by considering the difference between successive histograms,
that is,
\begin{align}
	\Vecy_i &= \Vecm_{i+1} - \Vecm_{i} \approx \left (\mv_{i+1}+\Vecb+\sum_{j=0}^{i}\Vecu_j \right)- \left (\mv_{i} + \Vecb + \sum_{j=0}^{i-1}\Vecu_j \right) \approx  \Vecu_i.
\end{align}
Due to the hemispherical reflection of light from a Lambertian ground plane and the occlusion effect of the vertical edge, the histogram differences %$\Vecy_i$ 
$\{\Vecy_i\}_{i=1,\dots,44}$ 
correspond to distinct wedges fanned out from the vertical edge.
We note that each photon detection time histogram $\Vecm_i$ has Poisson-distributed entries, so each histogram difference $\Vecy_i$
has entries following the Poisson-difference or Skellam distribution\cite{Skellam1946}.
Moreover, the entries of $\Vecy_i$ are conditionally independent given the scene configuration.
Note that although the mean visible scene and ambient light contributions are removed by 
this procedure,
they do still contribute to the
variance of the observation noise;
see Supplementary Section 2, which discusses how working with $\Vecy_i$ directly instead of $\Vecm_{i}$ leads to a more efficient reconstruction procedure.

%% file: subfiles/03_lightTransport.tex
Whereas active NLOS methods typically attempt full 3D reconstructions, and passive edge cameras form a 1D representation of the hidden scene,  ERTI produces 
an intermediate 
``2.5D'' representation, 
augmenting a 2D plan view (the positions and orientations of surfaces in the hidden space) 
with
the height of each surface. 
This representation is chosen as a compromise between the acquisition method,
which measures polar coordinate parameters (azimuth from the position with respect to the vertical edge and range from the time-resolved sensing),
and typical scenes, which are more naturally represented as being composed of planar facets than spherical or cylindrical shells.
Although there is no similarly direct mechanism to measure elevation angle, 
the duration of the temporal response from a scene patch contains information about its spatial extent and orientation.
To regularize the reconstruction problem, we use the fact that most commonplace
objects in the scene, such as humans, walls or furniture, present a base that starts from the ground plane. 
Using this assumption, the temporal response profile of a wedge then provides local height
information about the hidden scene.
Consequently, we suppose that hidden scenes
can be coarsely described by uniform-albedo vertical planar facets extending up from the ground plane.
Because considering histogram differences isolates the response from a single wedge, we can compute the response for each wedge of the hidden scene independently.
For indoor scenes, we also model the presence of a ceiling as a single additional surface, assumed parallel to the ground. 
Despite being a major exception to our vertical facet model,
inclusion of the ceiling component is necessary as it often reflects a significant amount of light.

\begin{figure}
    \centering
    {\phantomsubcaption\label{fig:modela}}
    {\phantomsubcaption\label{fig:modelb}}
    {\phantomsubcaption\label{fig:modelc}}
    {\phantomsubcaption\label{fig:modeld}}
    \includegraphics[trim={25mm 6.2cm 27mm 5.4cm},clip,width=0.8\linewidth]{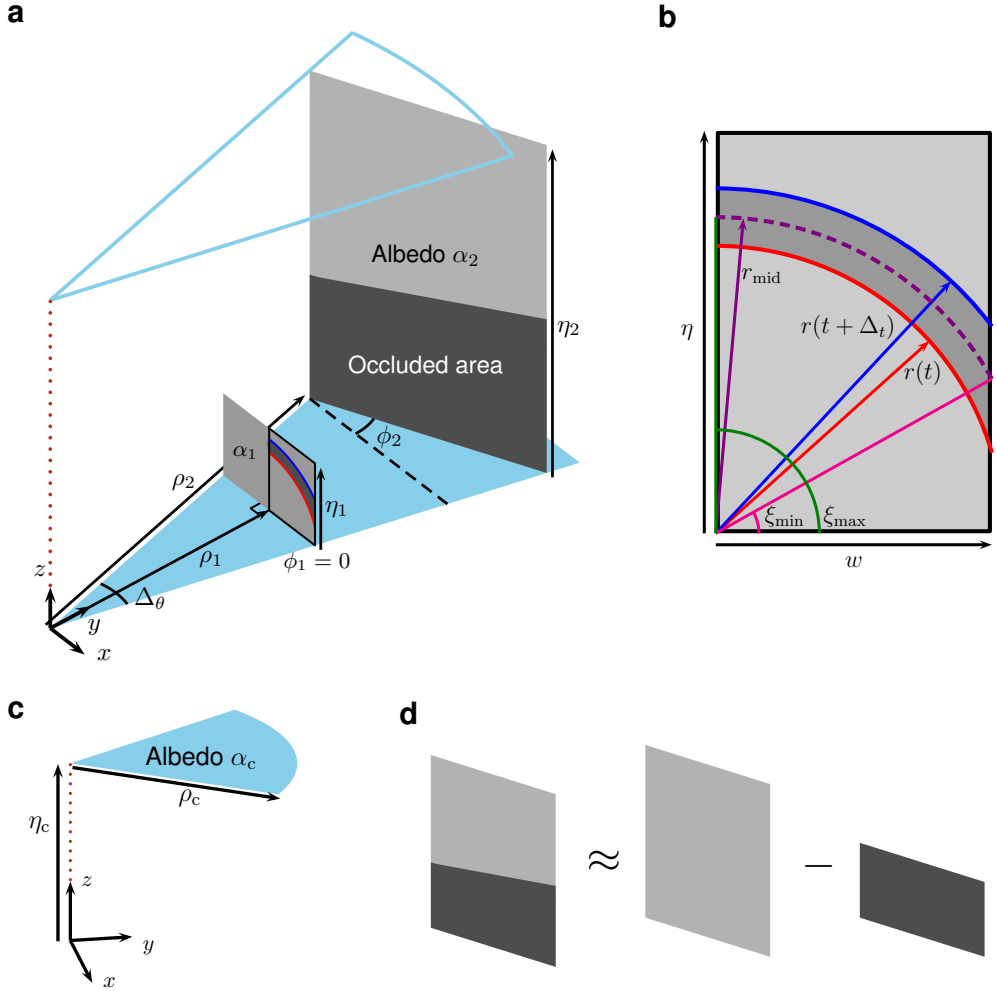}
    \caption{\textbf{Planar facet scene representation.} 
    \textbf{a}, The contents of a wedge spanning angle $\Dth$ are represented by a set of planar facets parameterized by a distance $\rho$, height $\eta$, albedo $\alpha$, and orientation angle $\phi$. 
    \textbf{b}, The basic transient light transport is computed for the region illuminated between times $t$ and $t+\Delt$ of one-half of a fronto-parallel facet. 
    For facets with nonzero $\phi$, the full response linearly combines two half-facet responses with the distance and widths adjusted according to orientation angle.
    \textbf{c}, The response from a ceiling component is computed similarly to a fronto-parallel facet. 
    \textbf{d}, Only the portion of a facet not occluded by a closer facet contributes to the total response.
    }
    \label{fig:representationDefs}
\end{figure}

The transient light transport for NLOS imaging with pulsed laser illumination and a focused detector is intricate but can be well approximated by factors accounting for the round-trip time of flight, radial falloff, and cosine-based corrections
for Lambertian reflection\cite{Heide2014,Thrampoulidis2017}.
Assuming the illumination arc radius, the SPAD FOV, and their separation are all small, 
the acquisition configuration is approximately confocal\cite{OToole2018}, with illumination and detection occurring at a single point at the base of the vertical edge. 
Without loss of generality and to simplify the light
transport model, this point is designated as the origin of the coordinate system both spatially and temporally -- the additional light travel time from the laser and back to the detector are subtracted away.
As shown in Fig.~\ref{fig:modela}, the wedge formed between illumination angles $\theta_{i+1}$ and $\theta_i$ is defined to have wedge angle $\Dth = \theta_{i+1}- \theta_{i}$.
Planar facets are parameterized by the shortest distance from the origin $\rho$, height $\eta$, albedo $\alpha$, and orientation $\phi$.
Because the ERTI azimuthal resolution is determined by the angular spacing of the illumination spots, we assume that planar facets span an entire angular wedge.

Using the confocal approximation, 
objects with the same path length to the origin lie on a sphere,
rather than a more general ellipsoid.
Over a time interval of duration $\Delt$ starting at time $t$ after the ground is illuminated, the intersection of the sphere with a planar facet is a section of a circular annulus.
We define the most basic transient response building block to be that for one-half of a fronto-parallel facet (i.e., $\phi=0$), as shown in Fig.~\ref{fig:modelb}, for which the circular sections are centred at a point on the ground plane at a distance of $\rho$ from the origin.
The full response of a fronto-parallel facet simply doubles the basic response due to symmetry, whereas the full response when $\phi\neq0$ adjusts the distance parameter based on the rotation angle and linearly combines two half-facet responses with different widths (see Supplementary Section~1).
For light with round-trip travel time $t$,
the radius of a circular section of a fronto-parallel facet is 
$r(t) = \sqrt{\left(\Frac{tc}{2}\right)^2-\rho^2}$.
The approximate angular limits of the annulus section as shown in Fig.~\ref{fig:modelb} are $\ximin = \cos^{-1} \left [ \min \{1, w/\rmid \} \right ]$ and $\ximax = \sin^{-1} \left [ \min \{1, \eta/\rmid \} \right ]$, which are defined with respect to the middle radius of the annulus $\rmid = \left [ r(t) + r(t+\Delt) \right ]/2$ and the half-facet width $w = \rho \tan(\Dth/2)$.
As detailed in 
Supplementary Section~1,
the transient light response for the annular section of a fronto-parallel facet thus approximately reduces to a computationally-efficient expression:
\begin{align} \label{eq:light_transport}
    g_t(\rho,\eta,\alpha,\Delt,\Dth)&\approx \frac{\alpha \rho^2}{12} [\ximax-\ximin+\sin(\ximin)\cos(\ximin) - \sin(\ximax)\cos(\ximax)] \nonumber \\
    & \qquad \times \left[\frac{3r^2(t)+\rho^2}{(r^2(t)+\rho^2)^3}-\frac{3r^2(t+\Delt)+\rho^2}{(r^2(t+\Delt)+\rho^2)^3} \right].
\end{align}
The combination of similar expressions for facets with nonzero orientation angle is likewise efficient to compute.
The transient response from an entire wedge also incorporates the portion of the ceiling within that wedge (Fig.~\ref{fig:modelc}), which also has a closed-form approximation similar to that of a fronto-parallel facet.
Finally, if multiple facets appear within a wedge, the total wedge response non-linearly combines facet contributions by removing the response components from more distant facets that are occluded by closer facets (Fig.~\ref{fig:modeld}).
The full derivation of the transient response for a wedge can be found in Supplementary Section~1. 

%% file: subfiles/04_reconstruction.tex
\begin{figure}[tp]
	\centering
	\includegraphics[width=\linewidth]{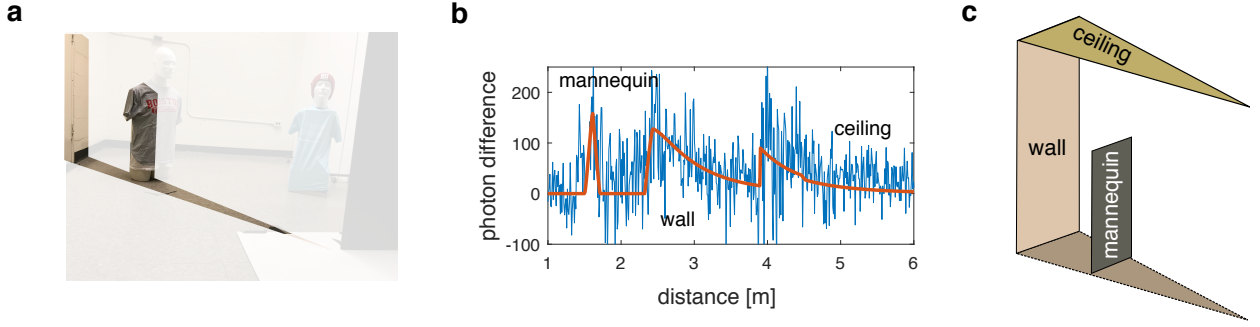}
	\caption{\textbf{Reconstruction of a hidden wedge.} 
	\textbf{a}, A photograph of an example hidden scene, highlighting the wedge to be reconstructed.
	\textbf{b}, The proposed algorithm fits the planar facet model (orange) to the acquired histogram difference (blue), identifying contributions from the mannequin, wall, and ceiling. 
	The time to each surface yields the position of the facets, the response shape provides information about the height and orientation, and the amplitude of the response is proportional to the surface albedo.
	\textbf{c}, This information is used to form a 2.5D reconstruction of the hidden wedge.}
	\label{fig:example_reconstruction}
\end{figure}

The reconstruction algorithm aims to fit the planar facet model to the observed histogram differences, as illustrated in \cref{fig:example_reconstruction}. 
One major difficulty of NLOS imaging is that the number of surfaces per resolved wedge is unknown a priori and can vary across the hidden scene.
Some wedges have only ceiling and wall contributions, whereas other wedges contain additional objects, such as the mannequins in our experiments. 
We simultaneously process all histogram differences to capture spatial dependencies between facet configurations across wedges of the hidden scene.

The performance of our reconstruction method relies on a carefully tailored scene model, which must be both flexible and informative while remaining computationally tractable. 
In natural hidden scenes, we observe that facets tend to be spatially clustered, 
with clusters representing different objects in the room.
We also observe that the positions of facets belonging to the same object tend to describe a 1D 
manifold. 
For example, the walls of the room can be described by a concatenation of facets 
forming the perimeter of the hidden scene.
Moreover, the parameters of neighbouring facets belonging to the same object are strongly correlated. 
For example, wall facets tend to share similar heights, albedos and orientations. 
These assumptions about scene structure % prior knowledge
are incorporated into the model via a Bayesian framework by defining a spatial point process prior model for the facet positions. 
This model is inspired by recent 3D reconstruction algorithms for LOS single-photon lidar that represent surfaces as 2D manifolds\cite{Tachella2018,Tachella2019}. 
Inference about the facet parameters 
(distance, height, albedo, and orientation angle)
and the ceiling parameters 
(height and albedo)
is carried out using a reversible-jump MCMC algorithm\cite{Green1995rjmcmc}, which 
maximizes a posterior distribution to find the most likely hidden room configuration given the observed data and prior beliefs (see Supplementary Section~2). 
At each iteration, the algorithm proposes a %stochastic 
random, yet guided,
modification to the configuration of facets (e.g., addition or removal of a facet), which is accepted 
with a pre-defined rule (the Green ratio\cite{Green1995rjmcmc}). Note that this approach only requires the local evaluation of the forward model, i.e., for individual wedges, which 
takes advantage of the fast calculations based on Equation~\eqref{eq:light_transport}. 
In particular, we can efficiently
take into account non-linear contributions due to occlusions between facets of a given wedge. 
By designing tailored updates (see Supplementary Section~2), 
the algorithm finds a good fit in few iterations, resulting in execution times of approximately 100~s, which is less than the acquisition time of the system.

%% file: subfiles/05_results.tex
Our reconstruction approach is 
assessed using measurements of
challenging indoor scenes containing multiple objects with a variety 
of depths, heights, rotation angles, and albedos.
The hidden scenes consist of an existing room structure modified by movable foamcore walls, with several 
objects placed within the scene.
Black foamboard is used to create a vertical edge with reduced initial laser reflection intensity.
Due to the specular nature of the existing floor, foamcore coated with a flat, white spray paint is used to achieve a more Lambertian illumination and detection relay surface, enabling even light distribution to all angles of the hidden scene.

\begin{figure}
    \centering
    \includegraphics[trim={25mm 73mm 25mm 61mm},clip, width=\linewidth]{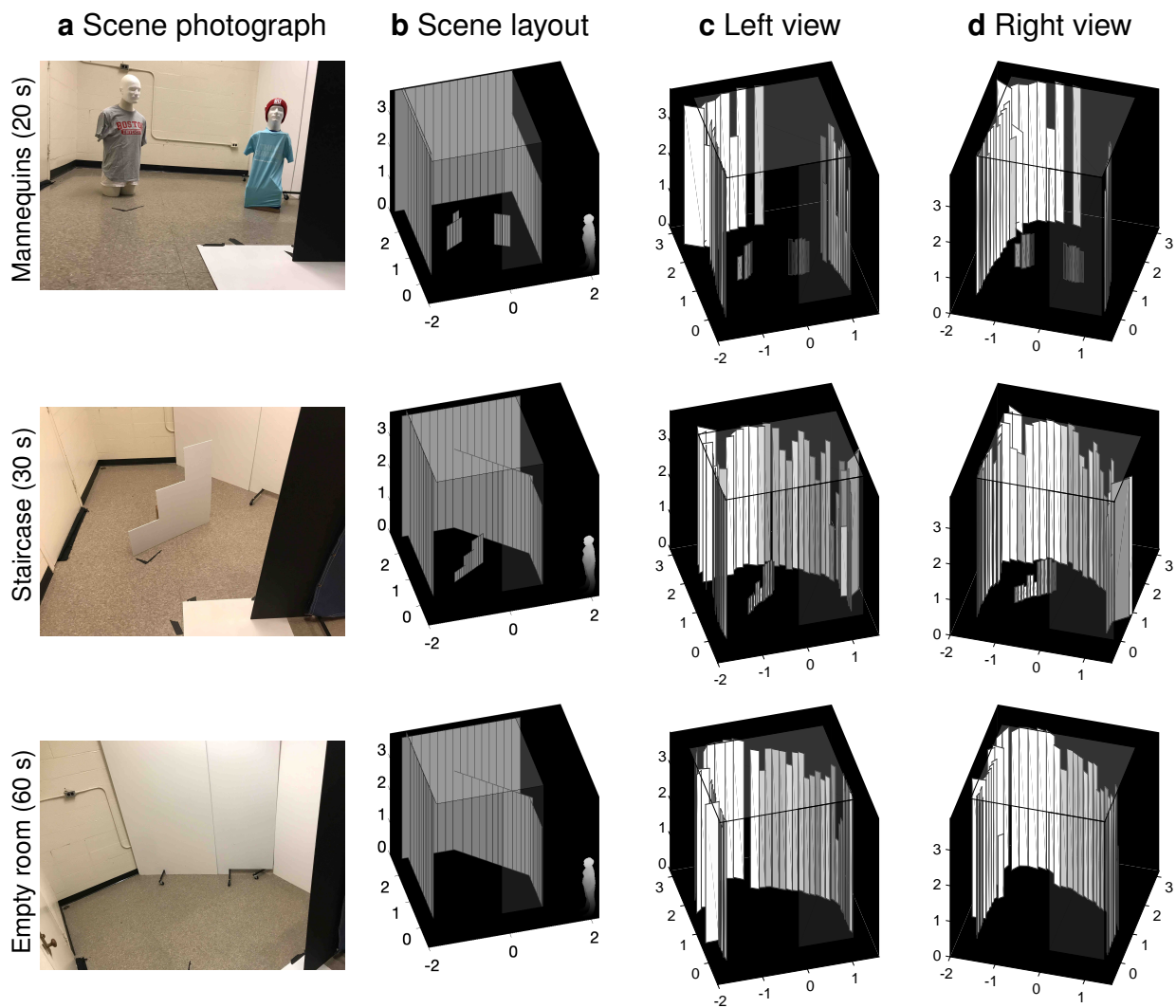}
    \caption{\textbf{Reconstructions of hidden scenes.} 
    Acquisition times are shown in seconds per illumination spot, and distance measurements are in meters.
    \textbf{a}, Hidden scene photographs show 
    rooms combining existing structures with moveable foamcore walls that are either empty or contain 
    two mannequins or a planar staircase.
    \textbf{b}, Diagrams show the approximate layout of the hidden scene plus the position of an observer. 
    \textbf{c}, Left \& \textbf{d}, right views of the reconstructed scenes.
    The foreground objects are correctly localized with height estimates within $\approx$10 cm, and most of the wall components are recovered.
    The estimated ceiling height and the known occluder edge are shown with partial transparency for context.}
    \label{fig:reconstructions}
\end{figure}

Fig.~\ref{fig:reconstructions} shows multiple views of the results of our reconstruction method for three example scenes. 
Each dataset was acquired from 45 illumination positions, with acquisition times of 20~s per illuminated spot for the mannequins, 30~s per illuminated spot for the staircase, and 60~s per illuminated spot for the empty room.
The approximate scene layout is displayed for reference using measurements from a laser distance meter.
The foreground objects, the ceiling height, and most of the wall components are recovered, with visual inspection confirming approximately correct positions and orientations.
The planar staircase object is useful as a height resolution test, with the average facet height for the 30-, 60-, and 90-cm steps measured to be 0.411, 0.543, and 0.921~m, respectively, yielding roughly 10-cm accuracy.
The most challenging components to accurately recover are wall facets that are occluded, oblique-angled, and/or far from the vertical edge. 
Additional results varying the scene content, acquisition duration, and number of illumination spots are presented in the Supplementary Section~6. 

In general, the histogram differences from real experimental data 
with reasonably short acquisition times
are extremely noisy (see Fig.~\ref{fig:example_reconstruction}), which makes accurate estimation challenging. 
Situations in which the visible scene response is large, or there is significant ambient background light, result in high variance in the measurements. 
Furthermore, the variance in the measurements due to the hidden scene itself increases linearly as a function of the illumination angle $\theta$, making the estimation more difficult at higher angles. 
Despite these effects, our reconstruction approach is quite robust to low signal strength and a high number of background counts, as confirmed by additional simulations presented in the Supplementary Section~7. 

%% file: subfiles/06_conclusions.tex
We have presented a method for imaging large-scale scenes outside the line of sight by measuring the transient light transport from scene illumination constrained by a visible occluder.
Other time-resolved methods for NLOS imaging using a relay wall have ellipsoidal uncertainty in the position of a reflecting surface, requiring a large scan area with many illumination and/or detection points.
The edge-resolving property of ERTI combined with the histogram differencing reduces the uncertainty from two dimensions to one, requiring dramatically fewer distinct measurements (e.g., 45 illumination locations) and a smaller aperture (e.g., 1.5~cm arc) than previous methods, as well as simplifying reconstruction.  
Moreover, existing
methods using the floor as a relay surface depend on differences between multiple acquisitions to isolate 2D positions of moving objects from the ``clutter’’ reflections from static surfaces\cite{Gariepy2016}, whereas ERTI recovers the entire static scene.

While we successfully demonstrate the ERTI acquisition and processing framework here, numerous aspects could be
improved through updated experimental and modeling approaches.
A straightforward method of decreasing the acquisition time would be to increase the laser power at the same wavelength\cite{Lindell2019a,Liu2019}.
Other works have even shown promising results with linear-mode avalanche photodiodes and lasers at higher wavelengths with greater eye-safety\cite{Brooks2019}.
Although we assume sequential illumination of evenly-spaced angles 
and a constant integration time for each spot, 
an alternative implementation could use a multi-resolution approach that first coarsely captures the hidden scene structure and then more finely samples areas that appear to have interesting content.
Finally, ERTI opportunistically uses a vertical edge to recover directional information from the hidden scene, but other more complicated occluder shapes could be used in conjunction with modified modeling.

%% file: subfiles/07_methods.tex
\begin{methods}

\subsection{Experimental Setup.}
A 120-mW master oscillator fiber amplifier (MOFA) picosecond laser (PicoQuant VisUV-532) at operating wavelength 532~nm is pulsed with repetition frequency $\fr=$ 20~MHz.
The illumination spot is redirected by a pair of galvo mirrors (Thorlabs GVS012), which is controlled by software through the analog outputs of a data acquisition (DAQ) interface (NI USB-6363).
Simultaneously with the illumination trigger, the laser sends a synchronization signal to the TCSPC electronics (PicoQuant HydraHarp 400), which starts a timer.
The ``stop'' signal for the timer is a detection event registered by the SPAD detector (Micro Photon Devices Fast-gated SPAD, photon detection efficiency $\approx$ 30\% at 532~nm).
These detection events may be due to true photon detections such as back-reflected signal or ambient light, or due to noise such as thermal dark counts or afterpulses.

The hardware is positioned approximately 2 m from the occluder edge.
The laser illuminates a set of $\nl$ spots $\{\ell_i \}_{i=1}^{\nl}$
along a semicircle of radius $\rl$ on the floor plane, with the vertical edge at the center.
The spots are linearly spaced in angle with $\ell_1$ at angle 0 completely occluded from the hidden scene, and $\ell_{\nl}$ at angle $\pi$ where none of the hidden scene is occluded.

The SPAD has a 25-mm lens mounted at the focal distance from the detector element, so that the SPAD field of view (FOV) is a small, approximately-circular spot of radius $\ro$ on the ground plane.
The SPAD is mounted on an articulating platform (Thorlabs SL20) and oriented  so that the center of the FOV is approximately co-linear with the intersection of the ground plane and the occluding wall, a distance $\rs\approx$~20~cm from the corner.
Mounted in front of the collection lens is a bandpass filter (Semrock MaxLine laser-line filter) with a transmission efficiency of $>90\%$ at the operating wavelength and a full width at half maximum (FWHM) bandwidth of 2~nm to reduce the amount of ambient light incident on the detector.
The timing offset of the laser/SPAD system is adjusted such that round-trip time of flight to and from the corner spot is removed (i.e., the corner point is at time zero).
Finally, a gate delay is adjusted so that the ``first-bounce'' light from the direct reflection is not recorded, to ensure that afterpulsing due to the strong direct reflection is minimized.
The SPAD gating is controlled by a delayer unit (MPD Picosecond Delayer) to have a gate-on duration of 42~ns starting $\approx 3$~ns after the peak of the direct reflection.

The laser is directed by the galvos to illuminate each spot in sequence for a time $\tdw$ per spot.
Detected photons are time-stamped by TCSPC module and streamed to the computer.
When the DAQ changes the galvo voltages to change the coordinates of the laser position, it simultaneously sends a marker to the TCSPC module 
indicating the spot to which the 
subsequent detections belong.
After the acquisition is completed, a histogram of detection times is formed for time bins with bin centers $\{b_i\}_{i=1}^{\nb}$, where $\nb = \lfloor \tr/\tbin \rfloor$ is the number of bins, $\tbin$ is the bin resolution, and $\tr = 1/\fr$ is the repetition period.
In this way, histograms can be formed for any histogram dwell time $\th$, where $\th\in[0,\tdw]$.

\end{methods}

%% file: SuppSubfiles/model_intro.tex
\section{Transient Light Transport Modeling}
\label{sec:TLTM}
The proposed method represents scenes primarily by a collection of vertical facets with four parameters: distance from the vertical edge $\rho$, height $\eta$, albedo $\alpha$, and orientation $\phi$.
While such a model is rich enough to describe many outdoor scenes,
the model also allows for indoor scenes, which usually have a ceiling, described as a horizontal planar facet parallel to the floor and with height $\eta_c$, maximum length $\rho_c$, and albedo $\alpha_c$.
This section demonstrates how those parameters are mapped to the continuous-time light intensity measured with an idealized pulsed illumination and time-resolved sensing system.
Also presented are the mappings for the discrete-time measurement systems used in practice, which integrate the light intensity over bins of duration $\Delta_t$.

By exploiting the inherent symmetry of diffuse light propagation, the full response for the intensity from a single wedge is based on the response for one half of a fronto-parallel facet ($\phi = 0$), which is derived in  Section~\ref{sec:frontoparallel}.
Section~\ref{sec:rotation} shows how the 
response for a facet with arbitrary orientation is derived by modifying the distance parameter and linearly combining the responses for two half-facets with different widths based on the orientation angle.
If multiple facets are present within a single wedge, closer facets will cause lower parts of the more distant facets to be occluded.
 Section~\ref{sec:occlusion} derives how the transient response from the occluded portion of the facet is subtracted off. 
 Section~\ref{sec:ceiling} derives an approximation to the response from the ceiling.
Finally,  Section~\ref{sec:fullResponse} shows how the facet and ceiling components are combined into the full wedge response.

%% file: SuppSubfiles/facetModel.tex
\subsection{Basic Transient Response Derivation for a Fronto-parallel Half-Facet}
\label{sec:frontoparallel}

Let $\Vecell$ be the position of a laser illumination, $\Vecc$ be a point in the
single-photon avalanche diode (SPAD)
field-of-view (FOV),
and $\Vecp$ be a point on a hidden surface $\CalS$.
Following previous models for non-line-of-sight (NLOS) imaging~\cite{Heide2014,Thrampoulidis2017},
the transient light transport due to illumination of $\Vecell$ at time $t_0$,
which has factors due to the round-trip time of flight, the radial falloff to and from the facet, and cosine factors for Lambertian reflection,
is given as
\begin{equation}\label{eq:genTransport}
    L(t) = \int_{-\infty}^t \int_{\CalS} \alpha(\Vecp) \frac{G(\Vecp,\Vecell,\Vecc)}{\norm{\Vecp-\Vecell}^2\norm{\Vecp-\Vecc}^2} \delta \! \left (t_0+(\norm{\Vecp-\Vecell}+\norm{\Vecp-\Vecc})/c-t' \right) \der \Vecp \der t',
\end{equation}
where $\alpha(\Vecp)$ is the albedo at point $\Vecp$,
$c$ is the speed of light,
the Lambertian bidirectional reflectance distribution function (BRDF) factor is
\begin{equation}\label{eq:Gdef}
    G(\Vecp,\Vecell,\Vecc) = \cos\left(\measuredangle(\Vecp-\Vecell,\nfl)\right) \cos\left(\measuredangle(\Vecell-\Vecp,\nfp)\right) \cos\left(\measuredangle(\Vecc-\Vecp,\nfp)\right) \cos\left(\measuredangle(\Vecp-\Vecc,\nfc)\right),
\end{equation}
$\nfl$, $\nfp$, and $\nfc$ are the normal vectors of $\Vecell$, $\Vecp$, and $\Vecc$, respectively, and $\measuredangle(\cdot,\cdot)$ denotes the angle between its vector arguments.

\begin{figure}
	\centering
	\includegraphics[width=0.7\linewidth]{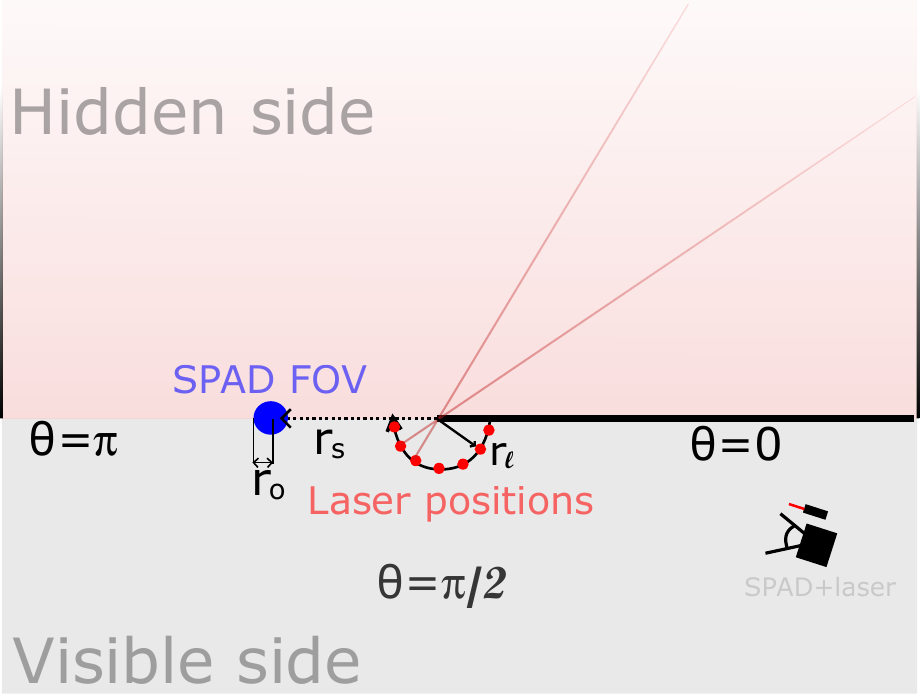}
	\caption{Top-down view of the imaging scenario.}
	\label{fig:topdown_angles}
\end{figure}

We assume the illumination arc radius $\rl$,
SPAD FOV radius $\ro$,
and the separation between illumination and SPAD FOV spots $\rs$ are small enough such that $\Vecell \approx \Vecc \approx \mathbf{0}$
(see Supplementary Figure~\ref{fig:topdown_angles}).
Then the acquisition configuration approximately corresponds to confocal illumination and detection~\cite{OToole2018} from a single point at the base of the vertical edge, which we define as the origin of our coordinate system.
The confocal configuration corresponds to a ``spherical geometry," such that objects for a given time of flight lie on the same sphere, rather than a more general ellipsoid.
We assume that the distance to the corner point is known, so we consider reflection off the floor at time $t_0=0$.
Equation~\eqref{eq:genTransport} thus simplifies to
\begin{equation}\label{eq:simpTransport}
    L(t) = \int_0^t \int_{\CalS} \alpha(\Vecp) \frac{G(\Vecp,\mathbf{0},\mathbf{0})}{\norm{\Vecp}^4} \delta \! \left (2\norm{\Vecp}/c-t' \right) \der \Vecp \der t'.
\end{equation}

We consider the light intensity contribution from a single wedge.
Our scene representation describes the world as being composed of Lambertian planar facets with a \emph{gravity prior} ensuring all planar facets begin at the ground plane.
Without loss of generality, we orient our coordinate system for a wedge so that the facet is centered in the $x$ direction on the $y$-axis. 
Our initial modeling is based on a fronto-parallel facet with 
unit normal vector $\nfp = [0,-1,0]$.
The unit normal vector of the ground is defined as $\nfg = [0,0,1]$.
In general, point $\Vecp$ on the facet has coordinates $[x,y,z]$.
Thus, we can simplify
\begin{align}
    G(\Vecp,\mathbf{0},\mathbf{0}) &= \cos\left(\measuredangle(\Vecp,\nfg)\right) \cos\left(\measuredangle(-\Vecp,\nfp)\right) \cos\left(\measuredangle(-\Vecp,\nfp)\right) \cos\left(\measuredangle(\Vecp,\nfg)\right) \nonumber \\
    &= \frac{(\Vecp \cdot \nfg) (-\Vecp \cdot \nfp) (-\Vecp \cdot \nfp) (\Vecp \cdot \nfg) }{\norm{\Vecp}^4} \nonumber \\
    &= \Frac{y^2 z^2}{\norm{\Vecp}^4},
\end{align}
where the dot $(\cdot)$ notation is used here to indicate an inner product.

\begin{figure}
	\centering
	{\phantomsubcaption\label{fig:FPdefa}}
    {\phantomsubcaption\label{fig:FPdefb}}
	\includegraphics[width=\linewidth]{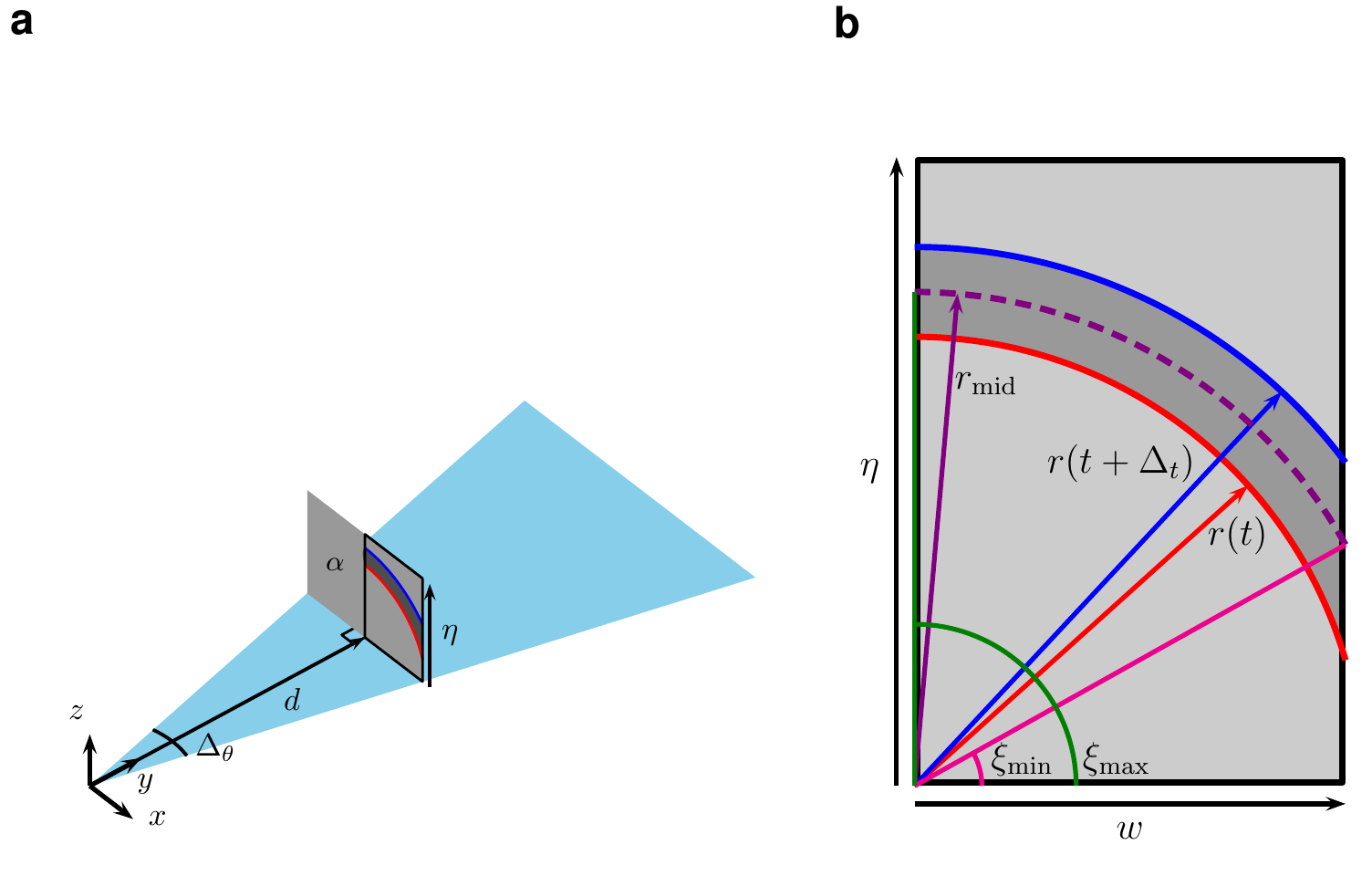}%{figs/facetDefs3}
	\caption{The basic transient response within a wedge is computed for one-half of a fronto-parallel planar facet with distance $d$, height $\eta$ and albedo $\alpha$.}
	\label{fig:sceneDef}
\end{figure}

The geometry for the fronto-parallel facet is shown in Supplementary Figure~\ref{fig:FPdefa}.
We define the perpendicular distance to the facet as $d$, so that the closest point to the origin is $[0,d,0]$, in the center of the bottom facet edge.
We define $\eta$ as the height and $\alpha$ as the uniform albedo of the facet that does not change with position.
We consider the transient response of the half-facet in the positive-$x$ half-plane, which has an angular span $\beta$.
The facet is assumed to span the width of the wedge regardless of the distance, so the half-facet width is $w = d \tan \beta$.
For the wedge between $\theta_{i+1}$ and $\theta_i$, define the wedge angle as $\Dth = \theta_{i+1}- \theta_{i}$, so
for the fronto-parallel facet we have $\beta = \Dth/2$.
Finally, define $h(t; d, \eta,\alpha,\beta)$ to be the transient response of the half-facet in the positive-$x$ half-plane.
Due to the horizontal symmetry of the facet, we can easily compute the full response of a fronto-parallel facet as
\begin{equation}
    \label{eq:full-fronto-parallel-from-half}
L(t) = 2h(t; d, \eta,\alpha,\beta).
\end{equation}
However, the half-facet response will also be useful for computing responses for arbitrary orientations as seen in Section~\ref{sec:rotation}. %the following subsection.
We can expand~\eqref{eq:simpTransport} to
\begin{align}
    L(t) &=  \alpha \int_0^t \int_{\CalS} \frac{y^2 z^2}{\norm{\Vecp}^8} \, \delta \! \left (2\norm{\Vecp}/c-t' \right) \der \Vecp \der t' \nonumber \\
    &=  \alpha  \int_0^t \int_x \int_y \int_z \frac{y^2 z^2}{(x^2+y^2+z^2)^4} \, \delta \! \left (\frac{2}{c}\sqrt{x^2+y^2+z^2}-t' \right) \der z \der y \der x \der t', \end{align}
so
\begin{equation}
    h(t; d, \eta,\alpha,\beta) =  \alpha d^2 \int_0^t \int_0^w \int_0^\eta \frac{z^2}{(x^2+d^2+z^2)^4} \, \delta \! \left (\frac{2}{c}\sqrt{x^2+d^2+z^2}-t' \right) \der z \der x \der t'
\end{equation}    
by introducing the integration limits from the half-facet dimensions (see Supplementary Figure~\ref{fig:FPdefb}).

The spherical propagation of the illumination from the origin intersects a fronto-parallel planar facet along a circle, hence it is convenient to change from Cartesian to cylindrical coordinates $(r,\xi,y)$, where $r^2 = x^2 + z^2$, $z = r \sin \xi$, $x = r \cos \xi$, and $\der z \der x = r \der r \der \xi$.
The challenge is in defining the regions of integration.
The set of valid radius values is $\CalR = \left \{r : 0 \leq r \leq \max\{w,\eta\} \right \}$.
For a given radial coordinate $r$, the angles integrated are $\Xi = \left \{ \xi:  \cos^{-1}\left ( \min \{1,\Frac{w}{r}\} \right ) \leq \xi \leq \sin^{-1}\left ( \min \{1,\Frac{\eta}{r}\} \right ) \right \}$,
giving the transient light transport as
\begin{align}
    h(t; d, \eta,\alpha,\beta) &= \alpha d^2 \int_0^t \int_\CalR \int_\Xi \frac{r^2 \sin^2 \xi }{(r^2+d^2)^4} \, \delta \! \left (\frac{2}{c}\sqrt{r^2+d^2}-t' \right) r \der \xi \der r \der t' \\
    &= \alpha d^2 \int_\CalR \frac{r^3 }{(r^2+d^2)^4} \int_\Xi \sin^2 \xi  \der \xi \int_0^t \delta \! \left (\frac{2}{c}\sqrt{r^2+d^2}-t' \right) \der t'  \der r \\
    &= \alpha d^2 \int_\CalR \frac{r^3 }{(r^2+d^2)^4} \int_\Xi \sin^2 \xi  \der \xi \, H \! \left(\frac{2}{c}\sqrt{r^2+d^2}-t \right) \der r,
\end{align}
where $H(t)$ is the Heaviside step function.

Practical systems cannot measure the instantaneous transient response. 
TCSPC systems, for instance, accumulate photon detections within small time intervals (e.g., bins of a histogram).
Let $h_t(d, \eta,\alpha,\beta,\Delt)$  be the integral of the transient response over a bin of duration $\Delt$, i.e.,
\begin{align}
    h_t(d, \eta,\alpha,\beta,\Delt) &= \int_t^{t+\Delt} h(t'; d, \eta,\alpha,\beta) \der t' \nonumber \\
    &= \int_t^{t+\Delt} \alpha d^2 \int_\CalR \frac{r^3 }{(r^2+d^2)^4} \int_\Xi \sin^2 \xi  \der \xi \, H \! \left(\frac{2}{c}\sqrt{r^2+d^2}-t' \right) \der r \der t'.
\end{align}
We can then use the time dependence of the radius $r(t) = \sqrt{(\Frac{ct}{2})^2-d^2}$.
Although the limits of the inner-most integral unfortunately depend on $r$, the dependence is mild if the time duration $\Delt$ is small. 
Thus we fix an approximate angular range to be
from
$\ximin = \cos^{-1}( \min \{1,\Frac{w}{\rmid}\} )$
to 
$\ximax = \sin^{-1}( \min \{1,\Frac{\eta}{\rmid}\} )$,
where
$\rmid = [r(t)+r(t+\Delt)]/2$,
as shown in Supplementary Figure~\ref{fig:FPdefb}.
% \JMBcomment{I changed the denominators of $\ximin$ and $\ximax$ to $\rmid$}
Thus,
\begin{align}
    h_t(d, \eta,\alpha,\beta,\Delt) &\approx \alpha d^2 \int_{r(t)}^{r(t+\Delt)} \frac{p^3 }{(p^2+d^2)^4} \mathbbm{1}\! \left \{\frac{2d}{c}\leq t \leq \frac{2}{c}\sqrt{d^2+\eta^2+w^2} \right \} \der p \int_{\ximin}^{\ximax} \sin^2 \xi  \der \xi,
\end{align}
where $\mathbbm{1}$ is the indicator function.
The first integral evaluates to 
\begin{align}
    \int_{r(t)}^{r(t+\Delt)}  \frac{p^3 }{(p^2+d^2)^4} &\mathbbm{1}\! \left \{\frac{2d}{c}\leq t \leq \frac{2}{c}\sqrt{d^2+\eta^2+w^2} \right \} \der p \nonumber \\
    &= \frac{1}{12} \left[\frac{3r^2(t)+d^2}{(r^2(t)+d^2)^3}-\frac{3r^2(t+\Delt)+d^2}{(r^2(t+\Delt)+d^2)^3} \right] \mathbbm{1}\! \left \{\frac{2d}{c}\leq t \leq \frac{2}{c}\sqrt{d^2+\eta^2+w^2} \right \},
\end{align}
and the second integral evaluates to 
\begin{equation}
    \int_{\ximin}^{\ximax} \sin^2 \xi  \der \xi = \frac{1}{2}[\ximax-\ximin+\sin(\ximin)\cos(\ximin) - \sin(\ximax)\cos(\ximax)].
\end{equation}
Thus, for each time bin $[t,t+\Delt]$, we can compute
\begin{align}
    h_t(d, \eta,\alpha,\beta,\Delt) &\approx \frac{\alpha d^2}{24} [\ximax-\ximin+\sin(\ximin)\cos(\ximin) - \sin(\ximax)\cos(\ximax)] \nonumber \\
    & \qquad \cdot \left[\frac{3r^2(t)+d^2}{(r^2(t)+d^2)^3}-\frac{3r^2(t+\Delt)+d^2}{(r^2(t+\Delt)+d^2)^3} \right] \mathbbm{1}\! \left \{\frac{2d}{c}\leq t \leq \frac{2}{c}\sqrt{d^2+\eta^2+w^2} \right \}.
    \label{eq:h_t_approx}
\end{align}

%% file: SuppSubfiles/rotateFacet.tex
\subsection{Response Generalization to Arbitrary Orientation}
\label{sec:rotation}

A facet is fronto-parallel when it is perpendicular to the
line defining the center of a wedge.
In Section~\ref{sec:frontoparallel},
we determined the response of a fronto-parallel facet by adding the identical responses of two
%canonical
fronto-parallel half-facets in \eqref{eq:full-fronto-parallel-from-half}.
For a facet that is not fronto-parallel,
we can easily determine the response by taking advantage of the same calculations for fronto-parallel half-facets, but combining them differently.
Observe that a facet that is not fronto-parallel within a wedge can still be seen as a sum or difference of fronto-parallel facets
by rotating the coordinate system to align with the position of the point at the base of an
extension of the planar facet where the normal vector points toward the origin.
If that point is within the wedge, we have a sum
(e.g., addition of half-facets of width $w_1$ and $w_2$ in
Supplementary Figure~\ref{fig:my_label}(a));
if that point is outside the wedge, we have a difference
(e.g., half-facet of width $w_2$ subtracted from a half-facet of width $w_1$ in
Supplementary Figure~\ref{fig:my_label}(b)).
The closest point on the facet is no longer simply given as the distance to the facet along the $y$-axis.
Instead, we %We thus 
define $\rho$ to be the shortest distance from the origin to the facet. 
The particular nearest point depends on the magnitude $\magphi$ relative to the wedge angular extent.
Specifically, 
if $\magphi<\Dth/2$ (see Supplementary Figure~\ref{fig:rota}), the closest point to the origin is along the base of the facet, whereas if $\magphi \geq \Dth/2$, the closest point to the origin is at one of the bottom corners of the facet (see Supplementary Figure~\ref{fig:rotb}).
This definition was chosen so that the transient response for facets with the same value of $\rho$ but different orientation angles $\phi$ would still start at the same time, which helps with the convergence of the MCMC sampler. 

\begin{figure}
    \centering
    {\phantomsubcaption\label{fig:rota}}
    {\phantomsubcaption\label{fig:rotb}}
    \includegraphics[width=0.92\linewidth]{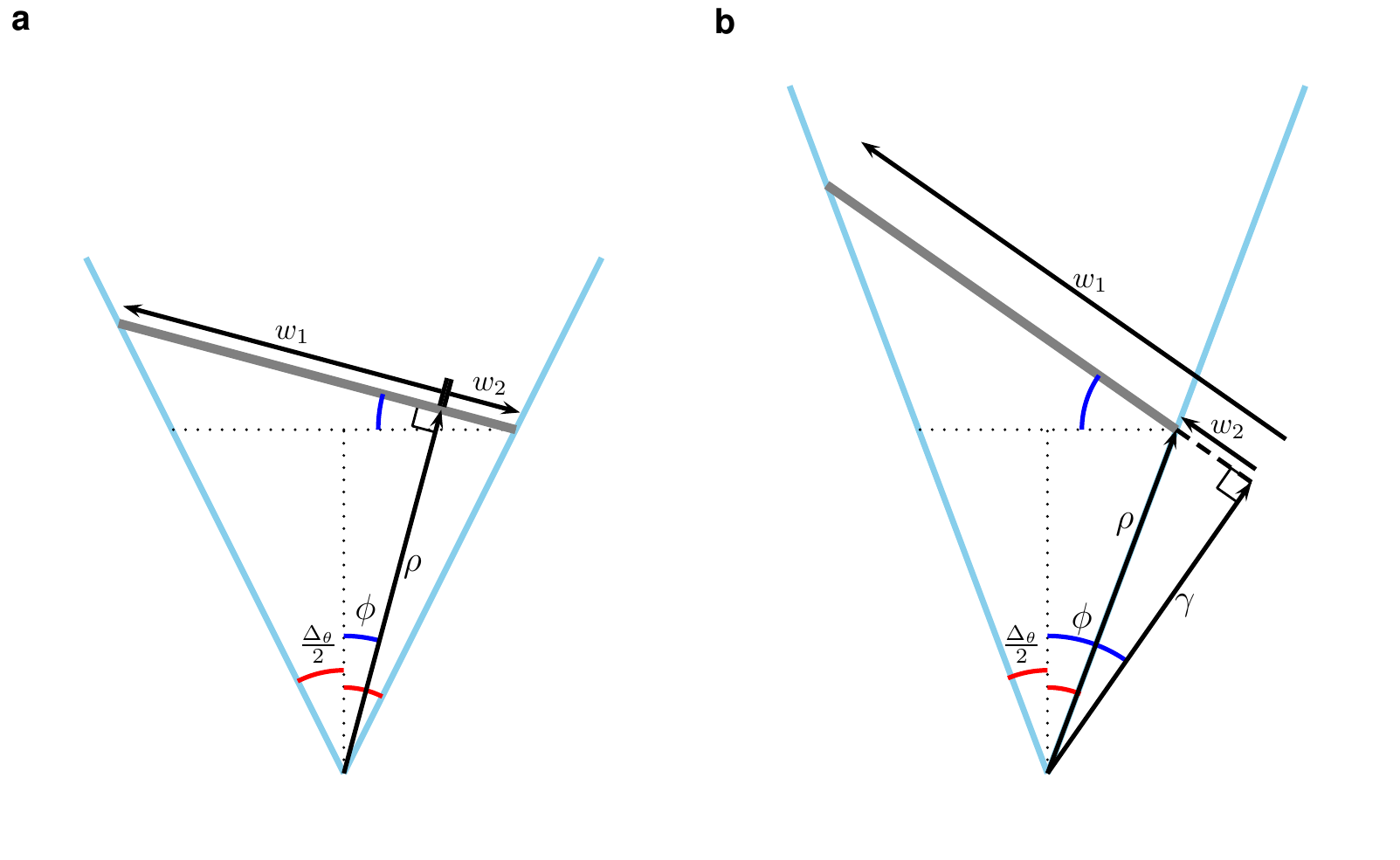}
    \caption{Top-down view showing parameter definitions for the facet response with rotation angle $\phi$.
    \subref{fig:rota}, If $\magphi<\Dth/2$, the responses for half-facets of lengths $w_1$ and $w_2$ are added. 
    \subref{fig:rotb}, Else if $\magphi>\Dth/2$, the response for the half-facet of lengths $w_2$ is subtracted from the response for the half-facet of lengths $w_1$. }
    \label{fig:my_label}
\end{figure}

The full response for a general facet linearly combines the responses for two fronto-parallel facets with different widths.
First, we determine the perpendicular distance to the facet to be $\gamma = \rho \cos(\max\{0,\magphi-\Dth/2\})$, which may be different from $\rho$ if $\magphi > \Dth/2$.
The angular spans of the two half-facets are
$\beta_1 = \left |\Dth/2-\magphi \right |$
and
$\beta_2 = \Dth/2+\magphi$,
yielding half-facet widths $w_1$ and $w_2$ corresponding to $w = \rho \tan \beta$.
Finally, the generic facet response is given as
\begin{align}
    g_t(\rho, \eta,\alpha,\phi,\Dth,\Delt) &= h_t(\gamma, \eta,\alpha,\beta_2,\Delt) %\nonumber \\
    %& \qquad
    + \text{sign}(\Dth/2-\magphi) \cdot h_t(\gamma, \eta,\alpha,\beta_1,\Delt).
\end{align}
Note that the full response from a fronto-parallel facet is a special case where $\phi=0$ and matches the expression found in the manuscript.

%% file: SuppSubfiles/occludedFacet.tex
\subsection{Incorporation of Occlusion Effect for Multiple Facets Within a Wedge}
\label{sec:occlusion}
\begin{figure}
    \centering
    \includegraphics[width=0.55\linewidth]{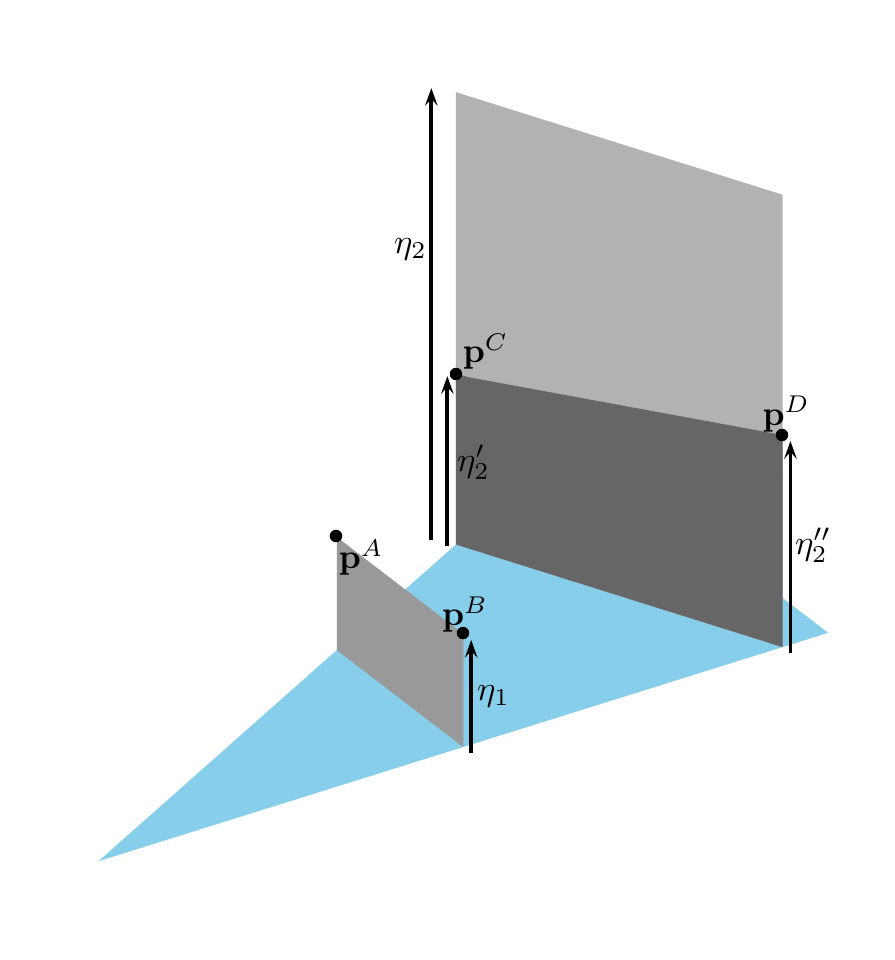}
    \caption{The response from a partially-occluded facet is computed by subtracting the approximate response from the occluded region. The occluded region approximation is determined by averaging the responses for facets of height $\eta_2'$ and $\eta_2''$.}
    \label{fig:compOcclReg}
\end{figure}

If multiple facets are present within a single wedge, then 
closer facets occlude the lower parts of farther facets, so the responses from multiple facets do not combine linearly.
To deal with occlusion, we compute the response from each facet in order of the proximity to the origin (i.e., sorted by $\rho$).
We then determine whether or how much of the later facets are visible, given the occlusion effects of closer facets.
For instance, Supplementary Figure~\ref{fig:compOcclReg} shows an example in which part of the second facet is still visible.
We thus compute the coordinates of the corners of the occluded region, approximate the response of the trapezoidal occluded region as the average of rectangular facet responses, and subtract the response of the occluded region from the total response.

Building on the previous section, we have defined the perpendicular distance to the facet as $\gamma$.
Unless the facet is fronto-parallel (i.e., $\phi=0$),
one of the facet bottom corners will be closer than the other.
The closer corner is at a radial distance
$r_1 = \Frac{\gamma}{\left [\cos \! \left(|\magphi-\Dth/2| \right)\right ]}$,
and the farther corner is at radial distance
$r_2 = \Frac{\gamma}{\left [\cos \! \left(|\magphi+\Dth/2| \right)\right ]}$.
Taking the case that the right corner is closer (i.e., $\phi > 0$),
the cylindrical coordinates ($\Vecp = [p_r,p_\theta,p_z$]) of the top corners of the first facet in Supplementary Figure~\ref{fig:compOcclReg} are given as 
\begin{equation}
    \Vecp^B =[r_1, \theta_{\rm R},\eta_1], \qquad \Vecp^A =[r_2, \theta_{\rm L},\eta_1],
\end{equation}
where $\theta_{\rm L} = \pi/2+\Dth/2$ and $\theta_{\rm R} = \pi/2-\Dth/2$ use the fact that the positive $y$-axis is assumed to bisect the wedge.

In general, the region of one vertical plane occluded by another vertical plane will be trapezoidal, unless the planes are parallel (e.g., $\phi_1 = \phi_2$).
We approximate the response from the occluded region as the average response for planes of heights $\eta_2'$ and $\eta_2''$,
which are the heights of the top corners of the trapezoid.
If both $\eta_2' < \eta_2$ and $\eta_2''< \eta_2$, then the response of the occluded plane can be subtracted from the total response for a plane of height $\eta$.
Else if $\eta_2' > \eta_2$ or $\eta_2''> \eta_2$, we assume the unoccluded region contributes negligibly to the response within that wedge.

We use the same approach as before to determine the radial coordinates $r^C$ and $r^D$ of the second facet.
The heights of $\Vecp^C$ and $\Vecp^D$ can be easily determined as
\begin{equation}
    \eta_2' = p_z^C = p_r^C \cdot \frac{p_z^A}{p_r^A}, \qquad \eta_2'' = p_z^D = p_r^D \cdot  \frac{p_z^B}{p_r^B}.
\end{equation}
Finally, the approximate transient light response from a partially-occluded facet is
\begin{equation}
    g_t(\rho_2, \eta_2,\alpha_2,\phi_2,\Dth,\Delt) - \frac{1}{2} \left [g_t(\rho_2, \eta_2',\alpha_2,\phi_2,\Dth,\Delt)  + g_t(\rho_2, \eta_2'',\alpha_2,\phi_2,\Dth,\Delt) \right].
\end{equation}

%% file: SuppSubfiles/ceilingModel.tex
\subsection{Ceiling Response Derivation}
\label{sec:ceiling}
A model of the world assuming only fronto-parallel planar facets leaves out the contribution from the ceiling, which is a strong exception to the gravity prior.
However, computation of the response from a section of illuminated ceiling is not so different to that of other planar facets and is outlined as follows.
We assume an entire wedge of the ceiling plane is illuminated as shown in Supplementary Figure~\ref{fig:ceiling} and define $\etac$ to be the constant height of the plane.
We also define $\alphac$ to be the uniform ceiling albedo and choose $\rhoc$ to be the maximum radial extent of the ceiling for the purpose of the derivation.\footnote{In practice, the ceiling response tends to decay quickly, so the value of $\rhoc$ is set automatically by the maximum time bin in our algorithm's implementation (see Section~\ref{sec:parameterization}).
We thus also ignore the possibility of a facet occluding part of the ceiling within a wedge.}
The ceiling response is then given as $c(t; \rhoc, \etac,\alphac,\Dth)$.

\begin{figure}
	\centering
	\includegraphics[width=0.75\linewidth]{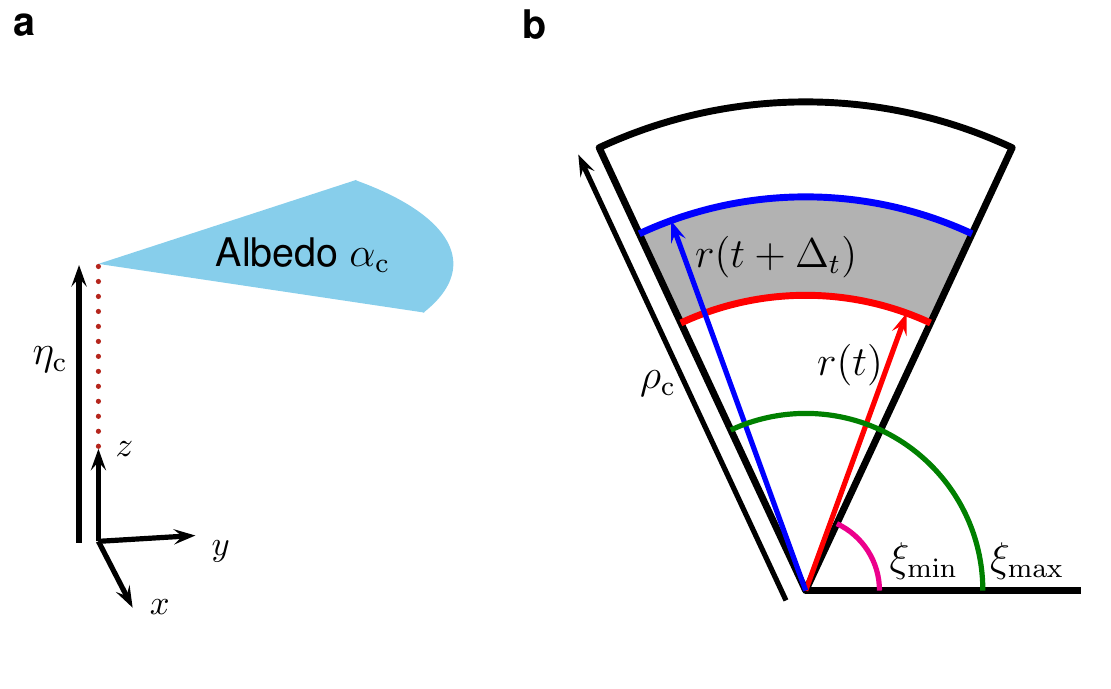}
	\caption{Ceiling contribution and forward modeling within a wedge.}
	\label{fig:ceiling}
\end{figure}

Once again, assuming a Lambertian ceiling, the temporal intensity response should be
given by \eqref{eq:simpTransport}.
The major difference is in the computation of the BRDF, since the floor and ceiling are assumed to be parallel to each other (i.e., no slanted ceiling).
The normal vector of any ceiling point is now $\nfp=[0,0,-1]$, so the BRDF factor is
\begin{align}
    G(\Vecp,\mathbf{0},\mathbf{0}) = \Frac{z^4}{\norm{\Vecp}^4}.
\end{align}
We can then once again expand~\eqref{eq:simpTransport} for the ceiling response: 
\begin{align}
    L(t) &=  \alphac \int_0^t \int_{\CalS} \frac{z^4}{\norm{\Vecp}^8} \, \delta \! \left( 2\norm{\Vecp}/c-t' \right) \der \Vecp \der t' \nonumber \\
    &=  \alphac  \int_0^t \int_x \int_y \int_z \frac{ z^4}{(x^2+y^2+z^2)^4} \, \delta \! \left( \frac{2}{c}\sqrt{x^2+y^2+z^2}-t' \right) \der z \der y \der x \der t', \end{align}
so
\begin{equation}
    c(t; \rhoc, \etac,\alphac,\Dth) =  \alphac \etac^4 \int_0^t \int_x \int_y \frac{1}{(x^2+y^2+\etac^2)^4} \, \delta \! \left( \frac{2}{c}\sqrt{x^2+y^2+\etac^2}-t' \right) \der y \der x \der t'.
\end{equation}    
We once again change to cylindrical coordinates with the longitudinal axis now aligned with the $z$-axis,
so $r^2 = x^2 + y^2$, $x=r\cos \theta$, $y=r\sin \theta$, and $\der x \der y = r \der r \der \theta$.
Assuming a maximum distance of the wedge $\rhoc$, 
we thus rewrite
\begin{equation}
    c(t; \rhoc, \etac,\alphac,\Dth) =  \alphac \etac^4 \int_0^t \int_0^{\Dth} \int_0^{\rhoc} \frac{r}{(r^2+\etac^2)^4} \, \delta \! \left( \frac{2}{c}\sqrt{r^2+\etac^2}-t' \right) \der r \der \theta \der t'.
\end{equation} 
Integrating over a time window $[t,t+\Delt)$ then produces
\begin{equation}
    c_t(\rhoc, \etac,\alphac,\Dth,\Delt) =  \alphac \etac^4 \Dth  \int_{r(t)}^{r(t+\Delt)} \frac{r}{(r^2+\etac^2)^4} \mathbbm{1}\! \left \{\frac{2\etac}{c}\leq t \leq \frac{2}{c}\sqrt{\etac^2+\rhoc^2} \right \} \der r.
\end{equation} 
The remaining integral has a closed-form expression,
\begin{equation*}
\int \frac{r}{(r^2+z^2)^4}\der r = \frac{-1}{6(r^2+z^2)^3},
\end{equation*}
so the measurement simplifies to 
\begin{equation}
c_t(\rhoc, \etac,\alphac,\Dth,\Delt)  = \frac{\alphac \etac^4 \Dth}{6} \left[\frac{1}{(r^2(t)+z^2)^3}-\frac{1}{(r^2(t+\Delt)+z^2)^3} \right] \mathbbm{1}\! \left \{\frac{2\etac}{c}\leq t \leq \frac{2}{c}\sqrt{\etac^2+\rhoc^2} \right \}.
\end{equation}

%% file: SuppSubfiles/wedgeModel.tex
\subsection{Full Wedge Transient Response Computation}
\label{sec:fullResponse}

The responses from the ceiling and $N_i$ facets in wedge $i$ are combined to compute the full wedge response $f_t(\rhoc,\etac,\alphac,\Vecrho,\Veceta,\Vecalpha,\Vecphi,\Dth,\Delt)$, where $\Vecrho$, $\Veceta$, $\Vecalpha$, and $\Vecphi$ are length-$N_i$ vectors.
If $N_i\geq 2$, the full wedge response computation includes the non-linear combination due to occlusion:
    \begin{align}\label{eq:fullResponse}
        f_t&(\rhoc,\etac,\alphac,\Vecrho,\Veceta,\Vecalpha,\Vecphi,\Dth,\Delt) \nonumber \\
        &=   c_t(\rhoc,\etac,\alphac,\Dth,\Delt)
           + g_t(\rho_1, \eta_1,\alpha_1,\phi_1,\Dth,\Delt) \nonumber \\
        & \quad + \sum_{k=2}^{N_i} \Big \{g_t(\rho_k, \eta_k,\alpha_k,\phi_k,\Dth,\Delt) 
           - \frac{1}{2} \left [g_t(\rho_k, \eta_k',\alpha_k,\phi_k,\Dth,\Delt)  + g_t(\rho_k, \eta_k'',\alpha_k,\phi_k,\Dth,\Delt) \right] \Big \},
    \end{align}
    which removes the occluded portion of a facet based on the height of the previous facet.
    It is assumed that the parameter vectors are ordered corresponding to $\rho_1 < \cdots < \rho_k < \cdots < \rho_{N_i}$.

%% file: SuppSubfiles/skellapop.tex
\section{Reconstruction Algorithm}

The reconstruction algorithm is a Bayesian method relying on reversible-jump Markov chain Monte Carlo (RJ-MCMC) simulation. It 
 requires (i) likelihoods for measurements given the set of scene parameters;
(ii) prior probability distributions for the scene parameters; and
(iii) a set of ``moves'' for generating samples in steps of the algorithm in order to explore the resulting posterior distribution of the set of scene parameters.
Scene parameterization is discussed in Section~\ref{sec:parameterization}.
The likelihoods are derived
in Section~\ref{sec:RJMCMC-likelihood}
from the response modeling in
Section~\ref{sec:TLTM}.
The prior model is based on natural relationships across wedges, such as continuity of surfaces, as described in Section~\ref{sec:RJMCMC-prior}.
The method for generating samples is described in Section~\ref{SEC:RJ-MCMC}. 
The full resulting algorithm is finally presented in Section~\ref{SUBSEC:full algo}.

\subsection{Parameterization}
\label{sec:parameterization}
The hidden room is characterized by a ceiling and a set of planar, vertical facets.
The ceiling contribution is defined by two parameters:
height $\ceilh$ and albedo $\ceila$, collectively referred to as $\ceil = [\ceilh,\ceila]^T$. 
The planar facet indexed by $n$ is described by the 4-tuple $(\Vecx_n,\intf{n},\hf{n},\af{n})$,
where $\Vecx_n=[\theta_n,\df{n}]^T$ is the position discretized by the $\nl$ illumination spots and $\nb$ time bins\footnote{The facet position in meters is obtained using the time bin duration $\Delt$ and wedge angle of the system $\Dth$.}, 
$\intf{n}\in\mathbb{R}_{+}$ is the (unnormalized) albedo,
$\hf{n}\in\mathbb{R}_{+}$ is the height, and
$\af{n}\in[0,\pi/2]$ the orientation angle.
The set of all planar facets across all wedges is defined as
\begin{equation}
    \pcloud = \{ (\Vecx_n,\intf{n},\hf{n},\af{n}), \quad n=1,\dots,\nfacets \}.
\end{equation}
Here we interpret the set of facets $\pcloud$ as a realization of a point process defined in 2D space $[1,\nl]\times[1,\nb]$,
where each point has additional \emph{marks} (i.e., properties) that are the height, albedo, and orientation of the facet.
Using a marked point process model, we do not fix a number $\nfacets$ a priori.
To simplify notations, we denote the set of point coordinates (without marks) as
$\pcloud_{\Vecx}=\{\Vecx_n,\, n=1,\dots,N_\Phi\}$.

\subsection{Likelihood}
\label{sec:RJMCMC-likelihood}
Assuming the incoming light flux incident on the single-photon detector is sufficiently low, the measured photon counts $m_{i,t}$ at bin $t$ for the $i$th illumination spot follow a Poisson distribution,
\begin{equation}\label{eq:poisson_model}
m_{i,t} \smid \Phi,\ceil,v_{t},b  \ssim \mathcal{P} \! \left (\sum_{j=1}^{i}u_{j,t} + v_{t}+ b \right),
\end{equation}
where $u_{j,t}=u_{j,t}(\Phi,\ceil)$ is the hidden scene component in the $j$th wedge and histogram bin $t$ that 
is approximately described by the
facet modeling and the ceiling parameters via \eqref{eq:fullResponse}, 
$v_{t}$ is the contribution from the visible side at bin $t$ (assumed to be the same across illumination spots), and $b$ is the background level due to ambient illumination. The histogram differences $y_{i,t} = m_{i+1,t}-m_{i,t}$ follow a Skellam\footnote{The Skellam probablity mass function is defined as
$P(x \smid \mu, \sigma) = \exp{(-\sigma^2)} \left( \frac{\sigma^2 + \mu}{\sigma^2 - \mu}\right)^{\Frac{x}{2}}  I_{x} (\sqrt{\sigma^4 - \mu^2})$, where $I_{x}$ the modified Bessel function of the first kind.} distribution~\cite{skellam1946frequency},
\begin{equation}\label{eq:skellam_lhood}
y_{i,t} \smid \Phi,\ceil \ssim \text{Skellam}(u_{i,t}, \sigma^2_{i,t}),
\end{equation}
with mean $u_{i,t}$ and variance
\begin{equation}
\sigma^2_{i,t}=u_{i,t} + 2\left(\sum_{j=1}^{i-1}u_{j,t} + v_{t}+ b\right).
\end{equation}
To improve the convergence properties of the MCMC algorithm and reduce its computational cost,
we have observed that it is preferable to work with the differences $y_{i,t}$ instead of the original measurements $m_{i,t}$.
This observation will be confirmed in Section~\ref{sec:model_comparison} by some simulation results.
Although the mean of the Skellam distribution in \eqref{eq:skellam_lhood} is only $u_{i,t}$ (i.e., only the wedge $i$), the variance $\sigma^2_{i,t}$ depends on the configuration of the scene in the previous wedges.
This prevents the model \eqref{eq:skellam_lhood} from being separable. 
Here, we first approximate the distribution of the measurements $y_{i,t}$ using composite marginal likelihoods~\cite{Varin2011composite}. 
More precisely, the joint likelihood of histogram differences $\DiffY \in \mathbb{Z}^{\nb\times\nl}$,
with $[\DiffY]_{i,t}=y_{i,t}$,
is approximated using separable factors
\begin{equation}
\label{eq:separable_model}
p(\DiffY \smid \Phi,\ceil)  \approx \prod_{i=1}^{\nb}\prod_{t=1}^{\nl}P(y_{i,t} \smid \Phi,\ceil),
\end{equation}
where $P(y_{i,t} \smid \Phi,\ceil)$ is the Skellam probability mass given by \eqref{eq:skellam_lhood}
and the variance is approximated by $\sigma^2_{i,t} \approx m_{i+1,t}+m_{i,t}$.
Using this first approximation, a hidden facet only affects the measurements linked to its wedge,
instead of the complete data, allowing local likelihood evaluation and parameter updates. Moreover, the estimation of the visible and background contributions can be bypassed.
Without loss of reconstruction performance, we further approximate the Skellam factors using their first two moments (i.e., a Gaussian density). This second approximation further reduces the computational cost associated with the likelihood evaluation.
An experimental comparison between the full likelihood and its separable approximations is presented in Section~\ref{sec:model_comparison},
validating the efficiency of these approximations.
Interestingly, we have empirically observed that the separable likelihood model \eqref{eq:separable_model} is more robust to imperfections of the sensing process
(e.g., non-ideal occlusion)
and mismatch between the recorded wedge response and the transient response model computed with \eqref{eq:fullResponse}.

\subsection{Prior Distributions}
\label{sec:RJMCMC-prior}
\subsubsection{Facet Positions}
\begin{figure}
	\centering
	\includegraphics[width=0.5\linewidth]{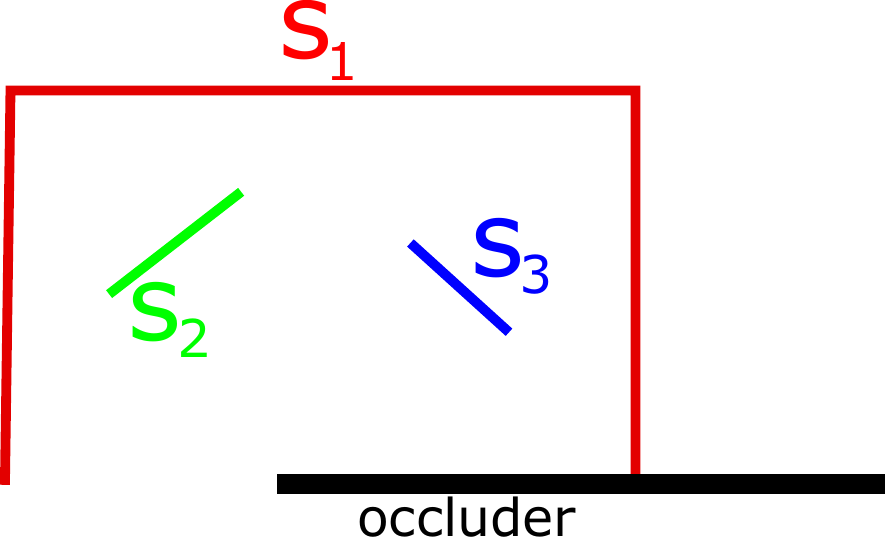}
	\caption{Top-down view of a hidden scene's 1D manifold structure. The facets of the room can be clustered together into 3 different 1D curves, i.e. $S_1$, $S_2$ and $S_3$. The position of the facets within each curve is highly correlated with its neighbours. We model these correlations with an area interaction process. Moreover, the different curves are usually clearly separated (in depth) within each wedge. This property is modeled with a repulsive Strauss process.}
	\label{fig:manifold_structure}
\end{figure}

We use a spatial point process prior distribution for the positions of the facets, similar to the one developed in the ManiPoP algorithm~\cite{Tachella2018}.
While ManiPoP defines a prior for 2D manifolds embedded in a 3D space, here we use similar ideas to model 1D manifolds in 2D space (wedge and depth coordinates).
This prior model is designed to promote spatial correlation between facets within the same object (e.g., a wall) and
repulsion between facets in the same wedge belonging to different objects (e.g., mannequin and wall in the same wedge).
Supplementary Figure~\ref{fig:manifold_structure} illustrates these phenomena.
The spatial point process prior is defined as a density $f$ with respect to a Poisson point process reference measure \cite[Chapter~9]{brooks2011handbook},
\begin{equation*}
f(\pcloud_{\Vecx} \smid d_{\min},\gamma_{a},\lambda_{a}) \propto f_1(\pcloud_{\Vecx} s \smid d_{\min}) f_2(\pcloud_{\Vecx} \smid \gamma_{a},\lambda_{a}),
\end{equation*}
where $f_1(\pcloud_{\Vecx} \smid d_{\min})$ and $f_2(\pcloud_{\Vecx} \smid \gamma_{a},\lambda_{a})$ are the Strauss and area interaction \cite{baddeley1995area} processes, respectively.
The repulsive Strauss process is defined as
\begin{equation*}
f_1(\pcloud_{\Vecx} \smid d_{\min}) \propto \left\{
\begin{array}{ll}
%0  & \text{if } \exists~ n \ne n' : x_{n}=x_{n'} \\
%&   \text{and } |\df{n}-\df{n'}| < d_{\min} \\
0, & \mbox{if there exists $n \ne n'$ such that $x_{n}=x_{n'}$ and $|\df{n}-\df{n'}| < d_{\min}$}; \\
1, & \mbox{otherwise},
\end{array}
\right. 
\end{equation*}
where $d_{\min}$ is the minimum separation in histogram bins between two facets in the same wedges. Attraction between points within the same surface is promoted by the area interaction process,
\begin{equation}
\label{EQ:area_interaction_density}
f_2(\pcloud_{\Vecx} \smid \gamma_{a},\lambda_{a}) \propto \lambda_{a}^{N_\Phi}\gamma_{a}^{-m\left( \bigcup_{n=1}^{\nfacets}S(\Vecx_n) \right)},
\end{equation}
where $m(\cdot)$ denotes the standard counting measure, $S(\Vecx_n)$ defines a set around the point $\Vecx_n$, and  $\gamma_{a}$ and $\lambda_{a}$ are two hyperparameters, accounting for the amount of attraction and total number of facets, respectively.
Both densities define Markovian interactions between points, only correlating points in a local neighbourhood (left and right wedges).
Moreover, the combination of both processes implicitly defines a connected-facet structure, which is used to model 1D manifolds in a (discrete) 2D space.

\subsubsection{Facet Parameters}
Since facets in neighbouring wedges usually correspond to the same hidden object (e.g., the walls, a piece of furniture, etc.),
their albedo, height and angle are generally correlated.
We model these correlations with Gaussian Markov random fields, making use of the implicit connected structure,
which only promotes correlation between facets linked to the same object.
To accommodate the compact support of the marks,
we introduce a change of variables that maps a compactly supported random variable into one defined on $\mathbb{R}$,
which is a standard procedure in spatial statistics~\cite{Rue2005}.
In our case, we define the mappings
%\begin{subequations}
\begin{align}
    \tintf{n} &= \log \intf{n}-\mu_{\intf{}}, \\
      \thf{n} &= \log \hf{n} -\mu_{\hf{}}, \\
      \taf{n} &= \log \left( \frac{\af{n}}{\frac{\pi}{2}-\af{n}} \right)-\mu_{\af{}},
\end{align}
%\end{subequations}
where $\mu_{\intf{}}$, $\mu_{\hf{}}$, and $\mu_{\af{}}$ are constants that center the distributions around the origin.
The prior distribution of the log-albedo is then defined by
\begin{equation} \label{EQ:a_joint}
\mathbf{\tintf{}} \smid \sigma_{\tintf{}}^2,\beta_{\tintf{}}, \rho, \pcloud_{\Vecx} \ssim \mathcal{N}(\mathbf{0},\sigma_{\tintf{}}^2\mathbf{P}^{-1}),
\end{equation}
where $\mathbf{\tintf{}} = [\tintf{1},\dots,\tintf{\nfacets}]^T$ is a vector containing all the intensities,
and $\sigma_{\tintf{}}^2$, $\beta_{\tintf{}}$, and $\rho$ are hyperparameters controlling the level of smoothness.
The precision matrix $\mathbf{P}$ is defined by
\begin{equation} \label{EQ:P}
[\mathbf{P}]_{n,n'}=
\begin{cases}
\beta + \sum_{\tilde{n}\in\mathcal{M}_{pp}(\Vecx_n)} \rho, & \text{if $n=n'$}; \\
      -\rho, & \text{if $\Vecx_n\in\mathcal{M}_{pp}(\cf{n'})$}; \\
          0, & \text{otherwise},
\end{cases}
\end{equation} 
where $\mathcal{M}_{pp}(\cf{n'})$ is the set of neighbours of point $\Vecx_n$, which is obtained using the connected-surface structure.
Note that the matrix $\mathbf{P}$ is very sparse due to the Markovian structure of the prior model.
Similar prior distributions are assigned to the transformed heights $\thf{}$ and angles $\taf{}$,
and they are defined by the hyperparameters $(\sigma_{\thf{}}^2,\beta_{\thf{}})$ and $(\sigma_{\taf{}}^2,\beta_{\taf{}})$, respectively.

\subsubsection{Ceiling Parameters}
We assign gamma prior distributions to the height and albedo of the ceiling,
\begin{align}
\ceilh \smid k_{\eta},  \theta_{\eta}   &\ssim \mathcal{G}(k_{\eta},\theta_{\eta}), \\
\ceila \smid k_{\alpha},\theta_{\alpha} &\ssim \mathcal{G}(k_{\alpha},\theta_{\alpha}),
\end{align}
where $(k_{\eta},\theta_{\eta})$ and $(k_{\alpha},\theta_{\alpha})$ are hyperparameters.
This choice is mainly motivated by the fact that it ensures the positivity of the height and albedo.

\subsection{Parameter Estimation} \label{SEC:RJ-MCMC}
Using the measurement likelihood from Section~\ref{sec:RJMCMC-likelihood}
and the priors from Section~\ref{sec:RJMCMC-prior},
the posterior distribution of the hidden room parameters (planar facets and ceiling) given the observations $\DiffY$ is
\begin{multline}
\label{EQ:posterior}
    p(\pcloud,\ceil \smid \DiffY, \Hyp )
    \ \propto \ p(\DiffY \smid \pcloud,\ceil) \,
                p(\tintf{} \smid \pcloud_{\Vecx},\sigma_{\tintf{}}^2,\beta_{\tintf{}}) \,
                p(\thf{} \smid \pcloud_{\Vecx},\sigma_{\thf{}}^2,\beta_{\thf{}}) \,
                \cdots \\
                p(\taf{} \smid \pcloud_{\Vecx},\sigma_{\taf{}}^2,\beta_{\taf{}}) \,
                f(\pcloud_{\Vecx} \smid d_{\min},\gamma_{a},\lambda_{a}) \,
                \pi(\pcloud_{\Vecx}) \,
                p(\ceilh \smid k_{\eta},\theta_{\eta}) \,
                p(\ceila \smid k_{\alpha},\theta_{\alpha}),
\end{multline}
where
$\Hyp=\{ \sigma_{\tintf{}}^2,\beta_{\tintf{}},\sigma_{\thf{}}^2,\beta_{\thf{}},\sigma_{\taf{}}^2,\beta_{\taf{}},k_{\eta},\theta_{\eta},k_{\alpha},\theta_{\alpha},d_{\min},\gamma_{a},\lambda_{a}\}$
defines the set of (fixed) hyperparameters. The hyperparameters have been set to
$$\Hyp=\{2.25,\, 11.25,\, 2.25,\, 11.25,\, 1,\, 5,\, 10,\, 0.5,\, 2,\, 1,\, 0.17c\Delta_t,\, e^4,\, e^{\Frac{\nl}{4}}\},$$
reflecting our prior knowledge about the unknown parameters
(e.g., the ceiling height should be approximately between 2 and 8 meters, the minimum distance between 2 facets within a wedge is 35 centimeters, etc.). 
 
Rather than the full posterior distribution in \eqref{EQ:posterior},
in this work, we compute the same posterior statistics as in \cite{Tachella2018}:
the facet parameters are estimated using the maximum a posteriori (MAP) estimator
\begin{equation}
\argmax_{\pcloud,\ceil}  p(\pcloud,\ceil \smid \DiffY,\Hyp),
\end{equation}
and the ceiling parameters are estimated using the minimum mean squared error estimator
\begin{equation}
\mathbb{E}\{\ceil \smid \DiffY,\Hyp\}.
\end{equation}
As these estimators cannot be derived analytically,
we  resort to Markov chain Monte Carlo (MCMC) simulation methods.
The simulation method, discussed in Section~\ref{sec:sampling},
generates $N_{m}$ samples of $(\pcloud,\ceil)$,
\begin{equation}
\{\pcloud^{(s)}, \ceil^{(s)}, \quad \mbox{for $s=0,1,\dots, N_{m}-1$}\},
\end{equation}
that are approximately drawn
from the posterior distribution \eqref{EQ:posterior}.
These samples are then used to approximate the estimators,
The estimate of the facet set is given by 
$\widehat{\pcloud} = \pcloud^{(\hat{s})}$, where
\begin{equation*}
\hat{s} = \argmax_{s \in \{0,...,N_m-1\}} p(\pcloud^{(s)},\ceil^{(s)} \smid \DiffY,\Hyp) .
\end{equation*}
The estimate of the ceiling is given by
\begin{equation*}
\hatceil = \frac{1}{N_{m}-N_{\textrm{bi}}}\sum_{s=N_{\textrm{bi}}+1}^{N_{m}}\ceil^{(s)},
\end{equation*}
where $N_{\textrm{bi}}$ is the number of burn-in iterations.

\subsubsection{Reversible-Jump Markov Chain Monte Carlo Sampling}
\label{sec:sampling}
As in ManiPoP~\cite{Tachella2018}, we use a reversible-jump MCMC algorithm that can handle the varying-dimension nature of the spatial point process
and allows us to build proposals tailored for the reconstruction problem;
see \cite{brooks2011handbook} for other stochastic simulation algorithms that can handle varying dimensions.
RJ-MCMC can be interpreted as a natural extension of the Metropolis-Hastings algorithm for problems with an a priori unknown dimensionality.
Due to the separable Skellam model and Markovian nature of the prior distributions,
all the proposed moves are local, having complexities proportional to the size of a single histogram of differences.
These moves are detailed in the following paragraphs. For ease of presentation, we summarize the main aspects of each move, referring the reader to~\cite{Tachella2018} for more a more detailed description of the different moves.

\paragraph{Birth and death moves.}
The birth move proposes a new point, sampled uniformly at random in the 2D space.
The albedo, height, and orientation angle are sampled from their prior distributions \eqref{EQ:a_joint}.
The complementary move, death, proposes the removal of a
facet chosen uniformly at random.%\VKG{\sout{ly chosen facet}}.

\paragraph{Dilation and erosion moves.}
Birth moves can suffer from low acceptance ratio, as the probability of randomly proposing a facet in the correct position can be low.
However, this problem can be overcome by using the current estimation of the surface to propose new facets in regions of high probability.
The dilation move proposes a point inside the neighbourhood of an existing surface with uniform probability across all possible neighbouring positions where a point can be added.
The rest of the parameters of the new facet (albedo, height, and orientation angle) are proposed in the same way as in the birth move
(sampled from their prior distributions).
The complementary move (named erosion) proposes to remove a point with one or more neighbours. 

\paragraph{Mark and shift moves.}
As in ManiPoP, the mark move updates the log-albedo of a randomly chosen point $\Vecx_n$. The albedo of the facet is updated independently using a Gaussian proposal with variance $\delta_{\tintf{}}$,
\begin{equation}\label{EQ:mark_prop}
\tintf{n}' \ssim \mathcal{N}\left(\tintf{n},\delta_{\tintf{}}\right).
\end{equation}
Similarly, the shift move updates the position of a uniformly chosen point within a wedge using a Gaussian proposal with variance $\delta_{\df{}}$,
\begin{equation}\label{EQ:shift_prop}
\df{n}' \ssim \mathcal{N}\left(\df{n},\delta_{\df{}}\right).
\end{equation}
The values of $\delta_{\tintf{}}$ and $\delta_{\df{}}$ are adjusted by cross-validation to yield an acceptance ratio close to $41\%$ for each move, which is the optimal value for a one dimensional Metropolis random walk, as explained in \cite[Chapter~4]{brooks2011handbook}. 

\paragraph{Sampling the ceiling parameters.}\label{SUBSEC:ceil sampling}
 The ceiling parameters are also sampled using a Metropolis random walk step, as in the mark and shift moves, i.e.,
\begin{align}\label{EQ:ceil_prop}
\ceil' \ssim \mathcal{N}\left(\ceil, \textrm{diag}(\delta_{\ceilh},\delta_{\ceila})\right).
\end{align}

\subsection{Full SkellaPoP Algorithms} \label{SUBSEC:full algo}
The RJ-MCMC algorithm chooses a different move at each iteration according to the probabilities in
Supplementary Table~\ref{TAB:RJMCMC params} and updates the ceiling parameters every $\nl\nb$ (total number of histogram bins) iterations.
The resulting algorithm, whose pseudocode is presented in Algorithm~\ref{ALG: SkellaPoP},
is referred to as SkellaPoP, as it handles measurements corrupted by Skellam noise and models the unknown parameters with a point process. 

\begin{table}
	\centering
	\begin{tabular}{| c | c | c | c | c | c|} % c | c|}
		\hline	$p_{\textrm{birth}}$ & $1/24$ & $p_{\textrm{death}}$ & $1/24$ & $p_{\textrm{dilation}}$  & $4/24$ \\  \hline $p_{\textrm{erosion}}$  &  $4/24$ &
		$p_{\textrm{shift}}$ &  $1/24$ & 	$p_{\textrm{mark}}$ &   $13/24$ \\ \hline % & 	
	\end{tabular}
	\caption{Move probabilities used in the RJ-MCMC sampler.}
	\label{TAB:RJMCMC params}
\end{table}

\begin{algorithm}
	\caption{SkellaPoP}\label{ALG: SkellaPoP}
	\begin{algorithmic}[1]
		\STATE \textbf{Input:} Histogram differences $\DiffY$, initial estimate 
		$(\pcloud^{(0)},\ceil^{(0)})$ and hyperparameters $\Hyp$
		\STATE \textbf{Initialization:} 
		\STATE $(\pcloud,\ceil) \gets (\pcloud^{(0)},\ceil^{(0)})$
		\STATE $s \gets 0$ 
		\STATE $\hatceil\gets 0$
		\STATE \textbf{Main loop:} 
		\WHILE{$s < N_{m}$}
		\IF {$\textrm{rem}(s,\nl\nb)==0$}
		\STATE  $(\ceil,\delta_{\textrm{map}}) \gets$ sample $\ceil$ using \eqref{EQ:ceil_prop}
		\ENDIF
		\STATE Choose move according to probabilities in Supplementary Table \ref{TAB:RJMCMC params}
		\STATE  $(\pcloud,\delta_{\textrm{map}}) \gets$ perform selected move 
		\STATE $\textrm{map} \gets \textrm{map} + \delta_{\textrm{map}}$
		\IF {$s \ge N_{\textrm{bi}}$}
		\STATE $\hatceil \gets \hatceil + \ceil$
		\IF {$\textrm{map}>\textrm{map}_{\max}$}
		\STATE $\widehat{\pcloud} \gets \pcloud$
		\STATE $\textrm{map}_{\max} \gets \textrm{map}$
		\ENDIF
		\ENDIF
		\STATE $s \gets s+1$
		\ENDWHILE
		\STATE $\hatceil \gets \hatceil/(N_{m}-N_{\textrm{bi}})$
		\STATE \textbf{Output:} Final estimates $(\widehat{\pcloud},\hatceil)$
	\end{algorithmic}
\end{algorithm}

To speed up the convergence of the RJ-MCMC algorithm,
we adopt a multiresolution approach in a fashion similar to \cite{Tachella2018}.
The dataset is downsampled by integrating photon detections in groups of 2 wedges.
Hence, the number of wedges is reduced by a factor of 2, leading to fewer points and background levels to infer with 2 times more photons per wedge.
The estimated point cloud at the coarse scale is upsampled using a simple nearest neighbour algorithm and used as initialization for the next (finer) scale.
In all our experiments we repeat the process for $K=2$ scales.
The multiresolution approach is finally summarized in Algorithm \ref{ALG: MR_approach}.
The ceiling parameters are initialized by first integrating the histogram differences across all wedges,
as all the ceiling contributions are approximately equal across wedges.
The first estimate of the ceiling's height is obtained by finding the maximum likelihood estimate of the integrated data,
in a similar fashion to matched filtering. 

\begin{algorithm}
	\caption{Multiresolution SkellaPoP}
	\begin{algorithmic}
		\STATE \textbf{Input:} Data $\DiffY$, hyperparameters $\Hyp$ and number of scales $K$.
		\STATE \textbf{Initialization:}
		\STATE $\pcloud_1^{(0)} \gets \emptyset$ (empty room)
		\STATE $\ceil_1^{(0)} \gets$ initialisation with integrated differences
		\STATE \textbf{Main loop:}
		\FOR{$k = 1,\dots,K$}
		\IF {$k>1$}
		\STATE $(\pcloud_k^{(0)},\ceil_k^{(0)}) \gets$ upsample$(\widehat{\pcloud}_{k-1},\hatceil_{k-1})$
		\ENDIF
		\STATE $(\widehat{\pcloud}_k,\hatceil_k) \gets$SkellaPoP$(\DiffY_k,\pcloud_k^{(0)},\ceil_k^{(0)},\Psi)$ 
		\ENDFOR
		\STATE \textbf{Output:}  $(\widehat{\pcloud}_K,\hatceil_K)$
	\end{algorithmic}
	\label{ALG: MR_approach}
\end{algorithm}

%% file: SuppSubfiles/comparison_observation_model.tex
\section{Observation Model Comparison}\label{sec:model_comparison}
We have compared the performance of the RJ-MCMC sampler using the original likelihood \eqref{eq:poisson_model} and the composite likelihood \eqref{eq:separable_model}, considering the following options:
\begin{enumerate}
    \item All scales Poisson: Correlated Poisson model \eqref{eq:poisson_model} at all scales. Note that this model is strictly equivalent to the Skellam-based model using difference histograms as input data and correlated variances.  
    \item Skellam: Separable composite likelihood with Skellam factors of variance $\sigma^2_{i,t} \approx m_{i+1,t}+m_{i,t}$ at all scales.
    \item Gaussian: Separable composite likelihood with Gaussian factors of variance $\sigma^2_{i,t} \approx m_{i+1,t}+m_{i,t}$ at all scales.
    \item Skellam denoised variance: Separable composite likelihood with Skellam terms of variance obtained by denoising $m_{i+1,t}+m_{i,t}$ under a temporal smoothness assumption.
    \item Gaussian denoised variance: Separable composite likelihood with Gaussian terms of variance obtained by denoising $m_{i+1,t}+m_{i,t}$ under a temporal smoothness assumption.
    \item Fine-scale Poisson: We run the sampler using the separable composite likelihood in the coarse scale and the correlated Poisson model in the fine scale. The coarse-scale estimate provides a good initialization for the fine scale, improving the mixing of the RJ-MCMC algorithm based on the correlated likelihood.
\end{enumerate}

\begin{figure}[th]
    \centering
    \includegraphics[width=1\textwidth]{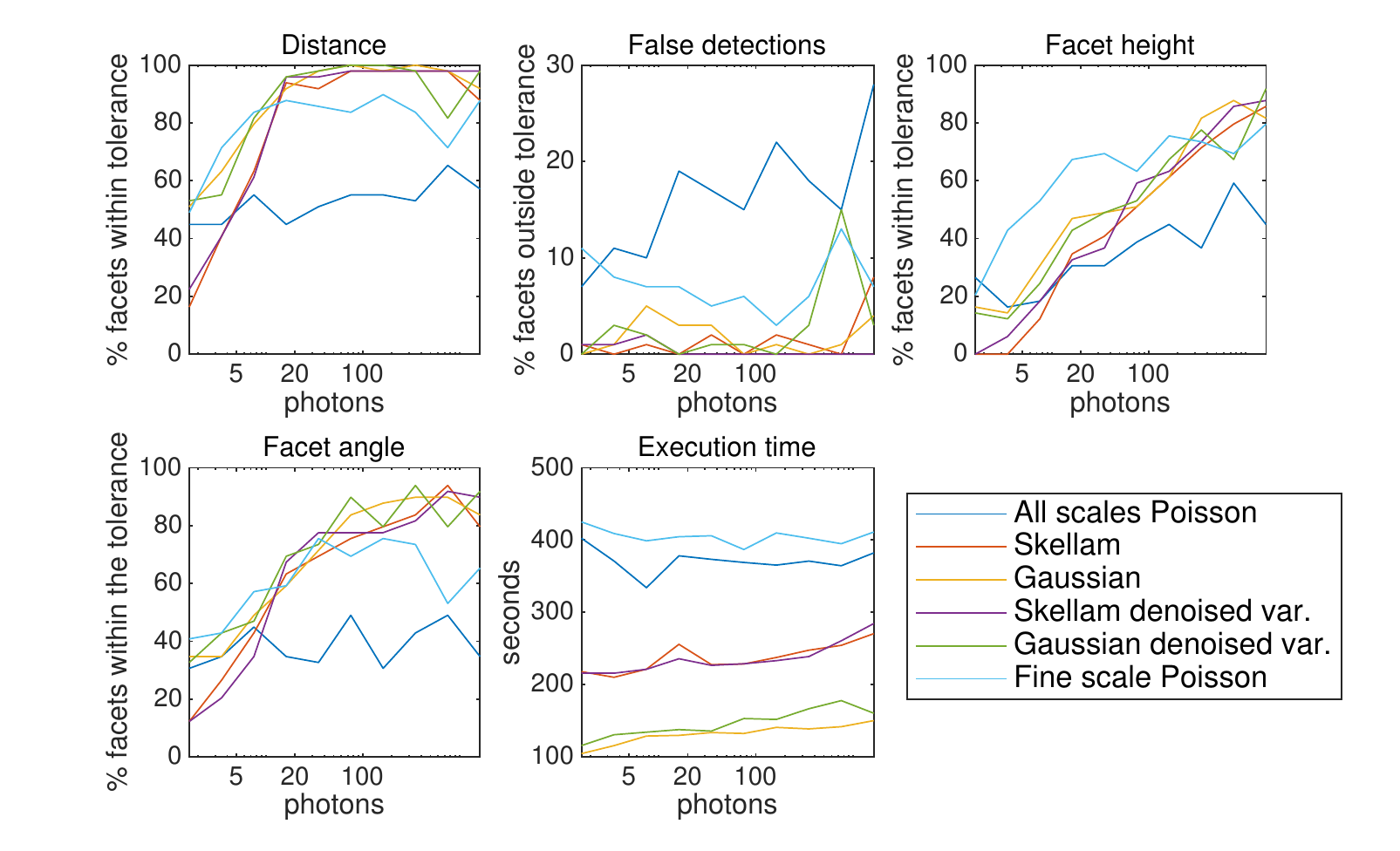}
    \caption{Impact of the observation model on the reconstruction performance. The results are shown for a tolerance of 20 cm for the distance, 50 cm for the facet height and 30 degrees for the orientation angle.}
    \label{fig:obs_model}
\end{figure}

We simulated a synthetic room using the full model \eqref{eq:poisson_model} to evaluate these six alternatives using the same number of scales and Monte Carlo iterations. The room is composed of 3 walls, a ceiling, and one central object, which partially occludes one of the walls. Supplementary Figure \ref{fig:obs_model} shows the performance of the evaluated algorithms, as a function of the mean number of recorded photons per histogram bin. 

The algorithms based on the full non-separable likelihood find more false detections and fail to find all the facets, even at high photon counts. This behaviour can be attributed to the bad mixing of the resulting sampler, which gets more easily stuck in a local maximum. Moreover, the execution time of these algorithms is significantly larger than that of the algorithms using composite likelihoods, as each likelihood evaluation (in the RJ-MCMC moves) requires a computation over the complete dataset of order $\mathcal{O}(\nl\nb)$. In contrast, the separable models only require a computation of order $\mathcal{O}(\nb)$. Both for the Skellam and Gaussian separable models, no significant improvement can be attributed to pre-estimating the variance with a denoiser. The Gaussian approximation performs better than the Skellam model, possibly due to the numerical instability of the numerical evaluation of the Skellam probability mass function. Moreover, the computation of the Skellam likelihood requires more operations, resulting in longer execution times than the Gaussian alternative. Hence, we have used the separable Gaussian model in the rest of our experiments.

%% file: SuppSubfiles/implementFigs.tex
\clearpage
\section{Additional Experimental Details}
Please see the Methods for a complete description of the experimental equipment.
The experimental setup used for data collection is shown in Supplementary~Figure~\ref{fig:expSetup}.
The FOV from the equipment location is shown in Supplementary~Figure~\ref{fig:expSetup2}.

\begin{figure}[hbtp]
    \centering
    \includegraphics[trim={0 0 7.8cm 0},clip,width=0.9\linewidth]{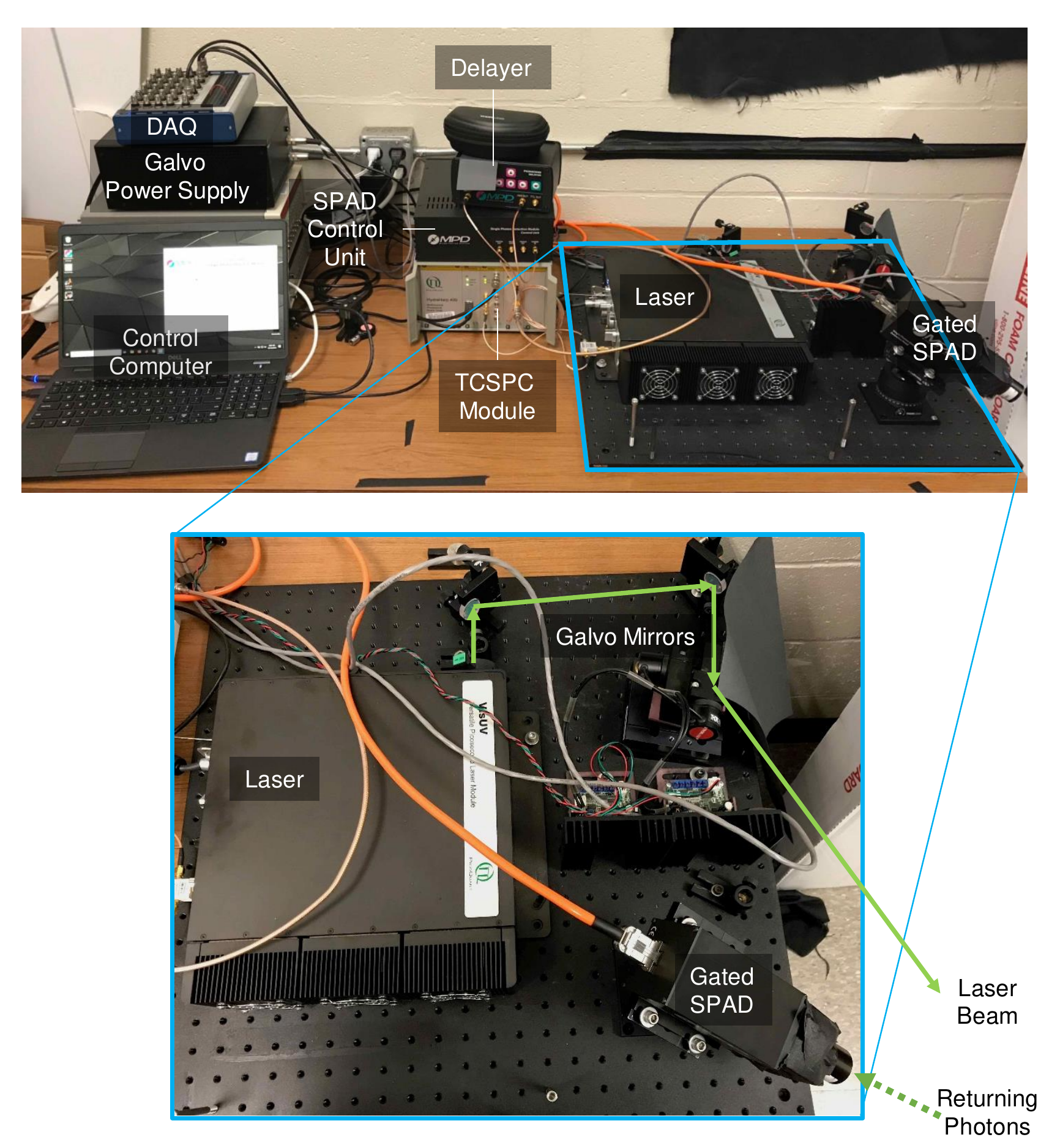}
    \caption{Equipment setup used for the data acquisition. See Methods for a detailed description of the acquisition procedure.}
    \label{fig:expSetup}
\end{figure}

\begin{figure}[t]
    \centering
    \begin{subfigure}[t]{0.49\linewidth}
        \centering
        \includegraphics[trim={0 0 11cm 0},clip,height=10cm]{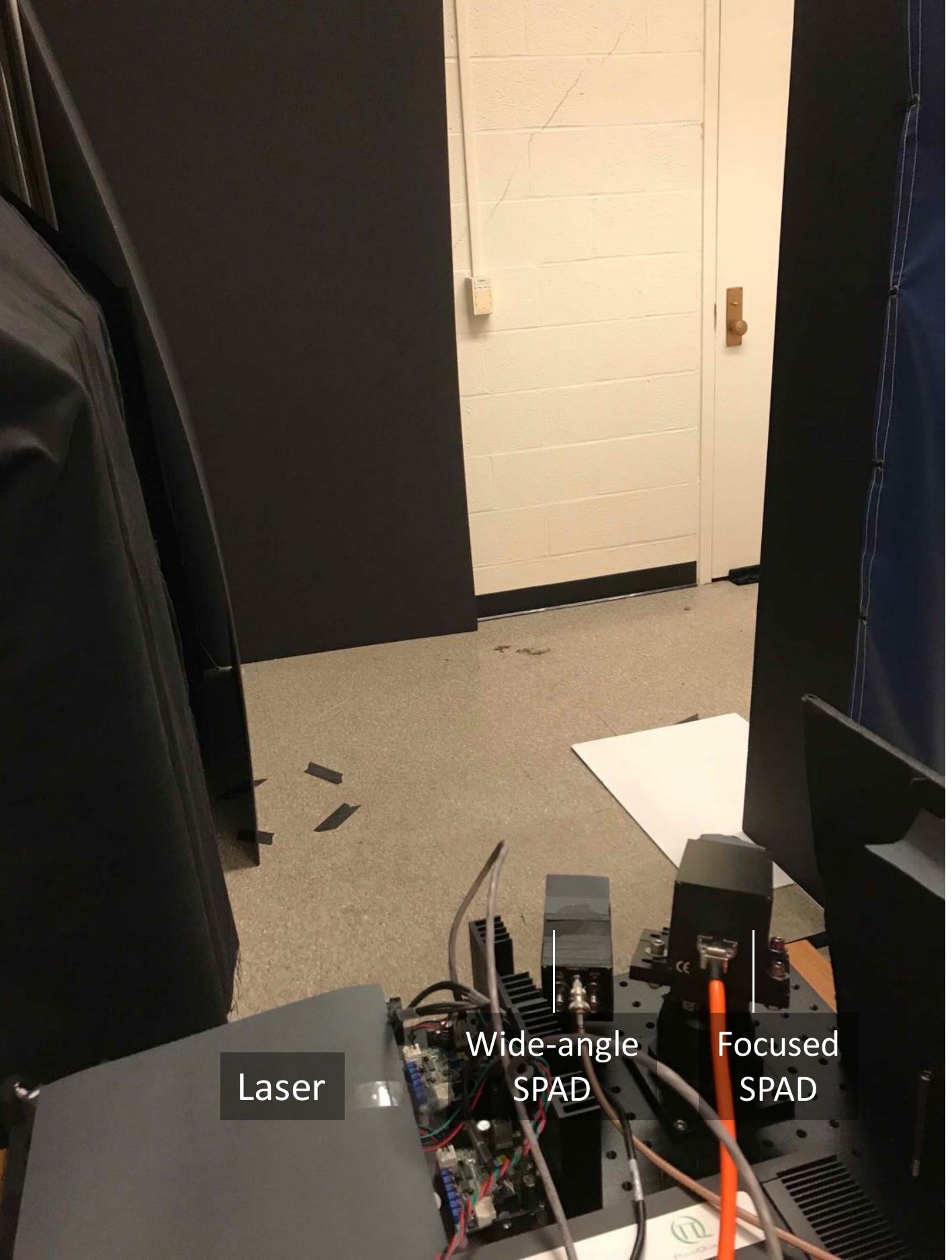}
        \caption{}
        \label{fig:setupView}
    \end{subfigure}
    \hfill
    \begin{subfigure}[t]{0.24\linewidth}
        \centering
        \includegraphics[trim={5mm 25mm 7mm 12mm},clip,height=10cm]{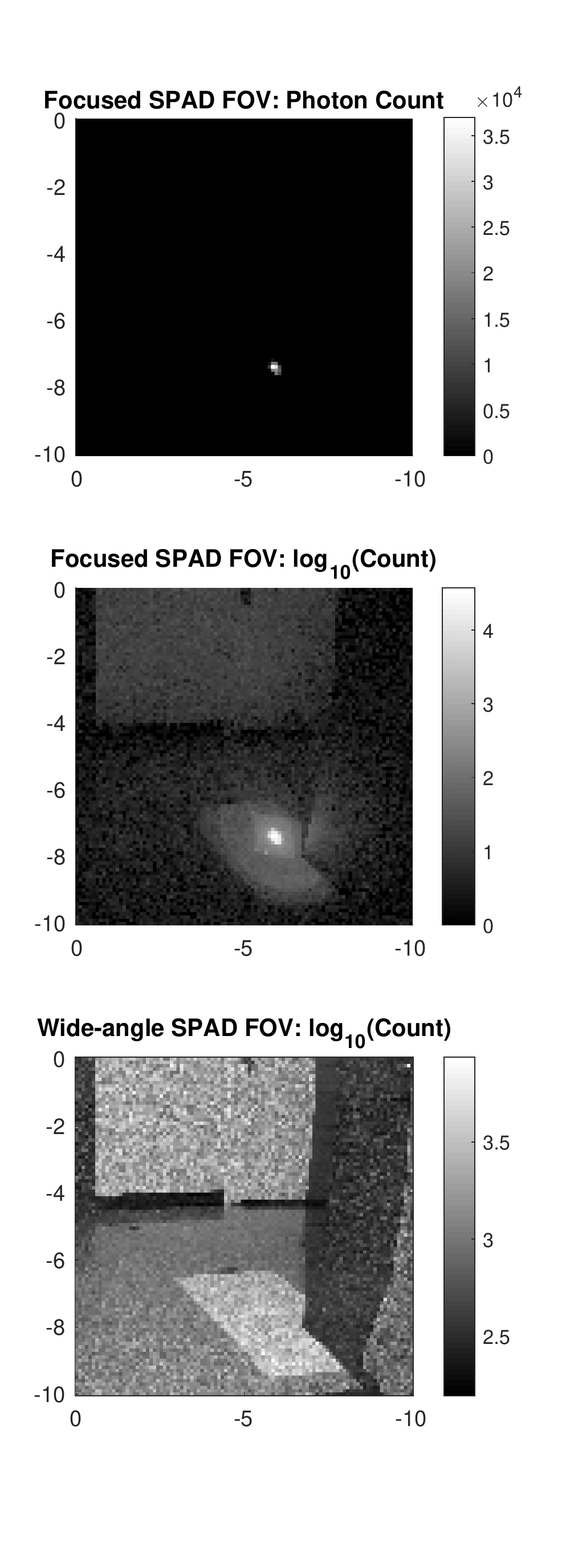}
        \caption{}
        \label{fig:spadFOV}
    \end{subfigure}
    \hfill
    \begin{subfigure}[t]{0.24\linewidth}
        \centering
        \includegraphics[height=10cm]{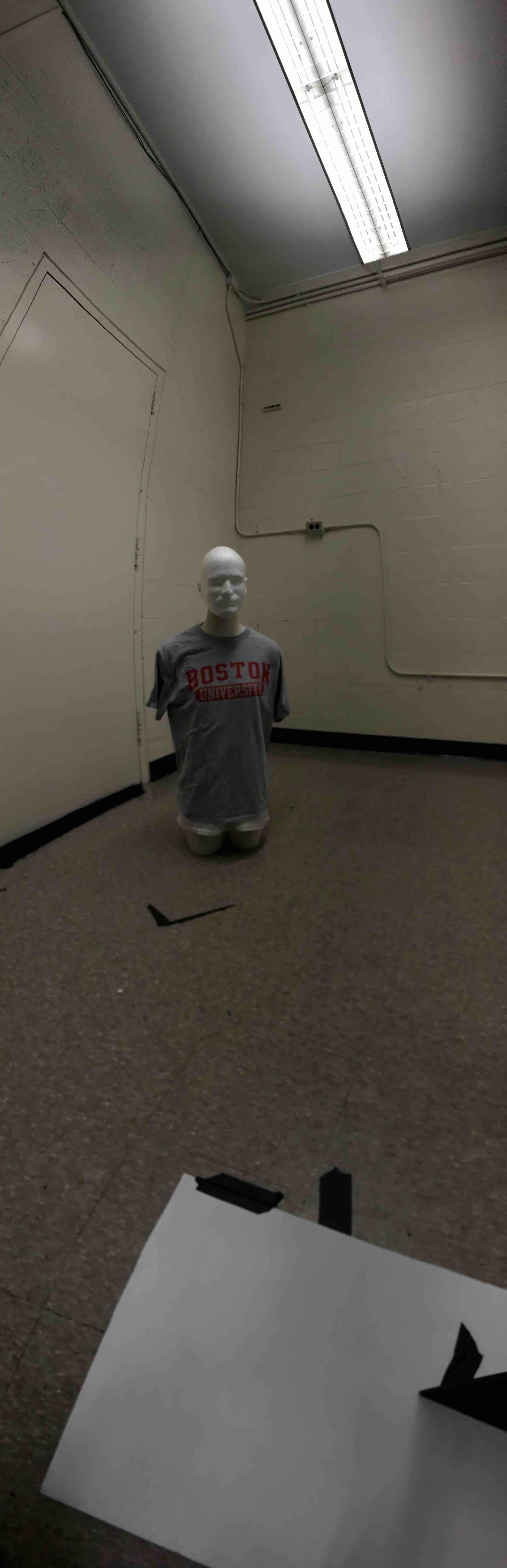}
        \caption{}
        \label{fig:sceneCeiling}
    \end{subfigure}
    \caption{Additional acquisition  details. \subref{fig:setupView} Scene visible from the acquisition equipment. The nearby walls in the visible scene are darkened to reduce the noise from the visible scene from limiting the ability to reconstruct the hidden scene.
    \subref{fig:spadFOV} The SPAD is focused onto a small spot approximately 20~cm beyond the vertical edge. In order to determine the SPAD FOV, the laser is scanned over a large area, and the number of returned photons is recorded for 10~ms at each position. A large number of photons is recorded only for when the laser is aimed within the SPAD FOV.
    Photon counts from laser positions outside the focused SPAD FOV are due to light that has undergone multiple reflections.
    A SPAD with a wide FOV can be used simultaneously to capture a more typical LOS image for reference. 
    \subref{fig:sceneCeiling} An extended view of the hidden scene. Note that the reconstruction method performs well despite the ceiling lamp and the slight glossiness of the walls not being included in the model. }
    \label{fig:expSetup2}
\end{figure}

%% file: SuppSubfiles/data.tex
\clearpage
\section{Example Data Set}

\begin{figure}[th]
    \centering
    \includegraphics[trim={0mm 35mm 0mm 30mm},clip,width=0.97\linewidth]{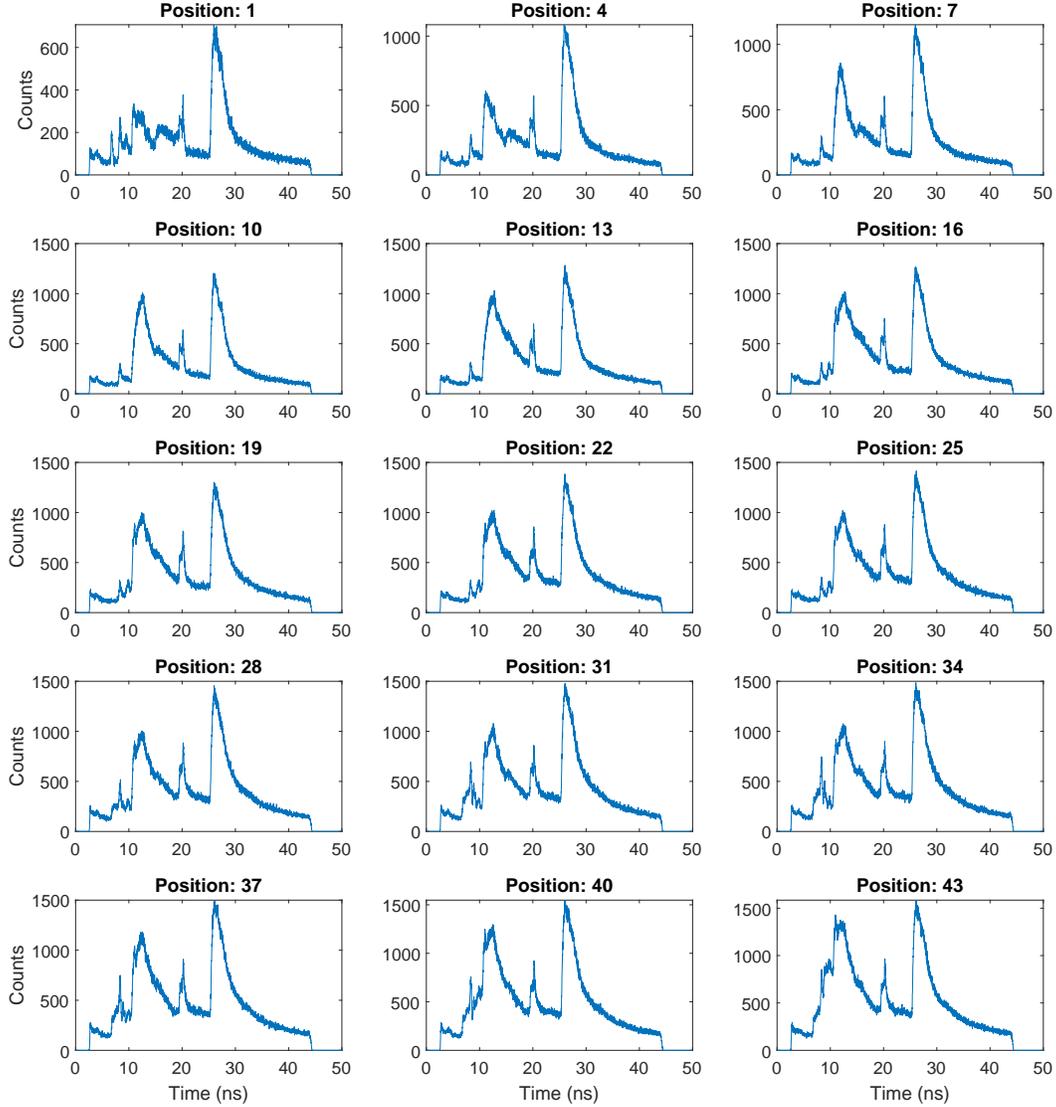}
    \caption{A subset of the measured data for the Mannequins (20 secs) scene in the first row of Figure 4 of the main paper.}
    \label{fig:data-hist}
\end{figure}
\begin{figure}[th]
    \centering
    \includegraphics[trim={0mm 35mm 0mm 31mm},clip,width=0.97\linewidth]{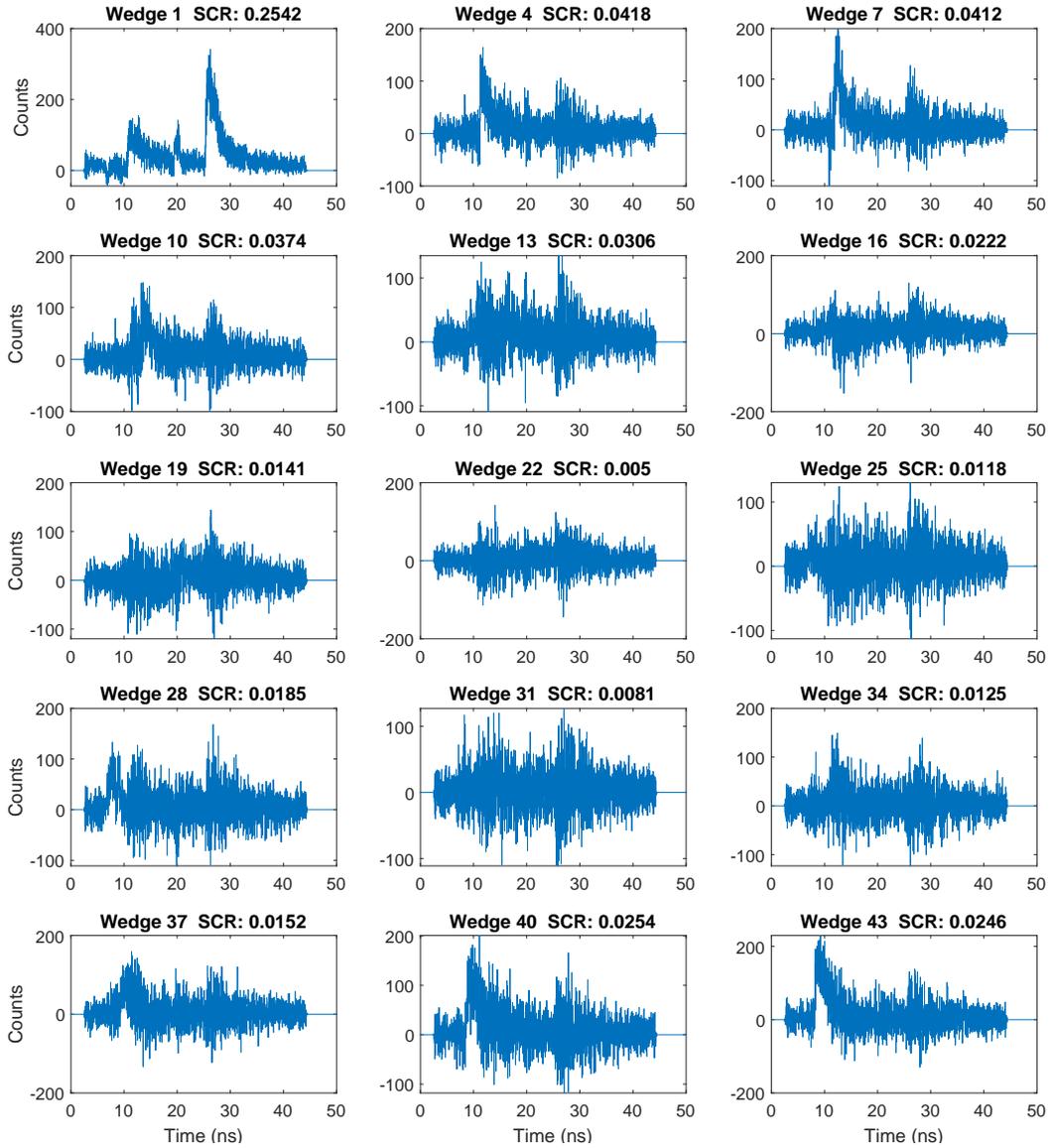}
    \caption{A subset of the differences between sequentially measured histograms for the Mannequins (20 secs) scene in the first row of Figure 4 of the main paper. Signal-to-clutter ratio (SCR) refers to the ratio of the mean counts in the histogram difference for wedge $i$ to the mean counts in the measurement histogram from position $i$.}
    \label{fig:data-hist-diff}
\end{figure}

Supplementary Figure~\ref{fig:data-hist} displays a subset of the directly measured photon detection time histograms used to form the reconstruction in the first row of Figure 4 (Mannequins, 20 secs). 
Supplementary Figure~\ref{fig:data-hist-diff} shows a subset of the differences between sequentially measured histograms, corresponding to the light detected from individual wedges.
Note that the histogram differences are far noisier than the histograms themselves. 
Included in the title of each subplot is the estimated \emph{signal-to-clutter ratio} (SCR), which we define to be the ratio of the mean counts in histogram difference $i$ to the mean counts in histogram $i$, containing all the counts from background, the visible scene, and previous wedges within the hidden scene. 
This value gives some indication of how much new information about wedge $i$ is contained in histogram $i+1$.

%% file: SuppSubfiles/experiments.tex
\clearpage
\section{Additional Results}
\begin{figure}[ht]
    \centering
    \includegraphics[trim={25mm 73mm 25mm 61mm},clip,width=\linewidth]{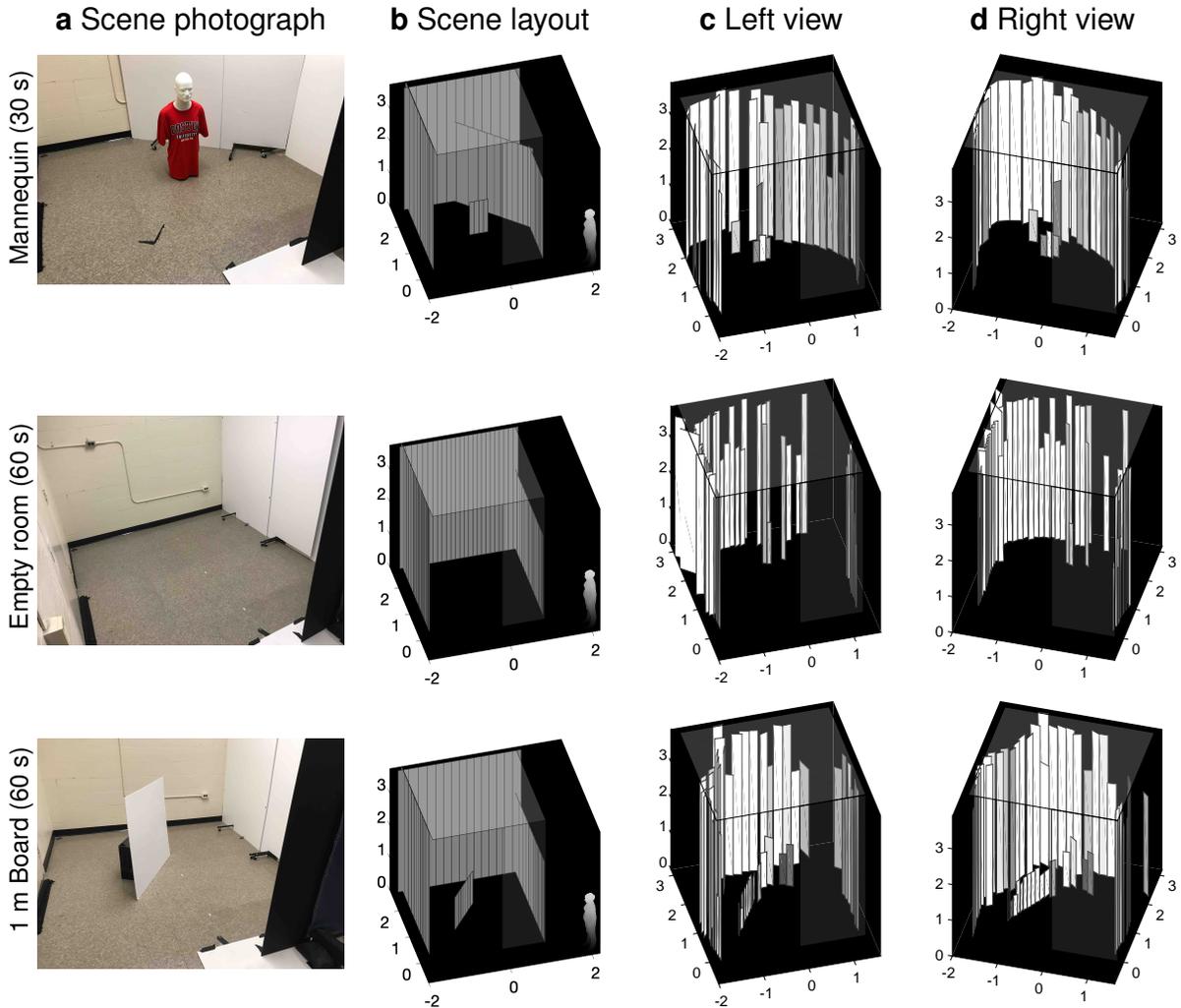}
    \caption{Reconstructions from additional acquisitions reinforce the performance of the acquisition and algorithmic approaches.}
    \label{fig:SupplResults}
\end{figure}

The results shown in Supplementary Figure~\ref{fig:SupplResults} depict additional scenes acquired with the same general procedure described in the Methods, but with some some variations in the imaging subject, acquisition time per spot, and numbers of illumination spots.
The top row shows the recovery of a scene from 37 illumination spots and with a challenging target: a mannequin wearing a red shirt that reflects very little of the green laser light.
The middle row shows a large, empty room recovered from 73 illumination positions.
The bottom row recovers the correct angle, position, and height of a 1~m $\times$ 1~m square planar target from 45 illuminations spots.

%% file: SuppSubfiles/robustness.tex
\section{Robustness to Signal Strength, and Ambient Light or Visible Scene Contributions}

Both ambient background light and visible scene contributions increase estimation difficulty. Although the mean contributions from both the background and visible scene cancel out due to the differencing of histograms, they still contribute to the variance of the measurements. Ambient background light contributions are assumed to be constant in time and so provide a constant background level within the measurement histograms. Visible scene light is time-varying and results in contributions similar to those of the visible scene, but that are approximately constant within each measurement histogram. This results in increased variance at different times within the measurements, making facets in the hidden scene, at distances similar to surfaces in the visible scene, more difficult to estimate. Similarly, the contributions from the ceiling, although modelled in our reconstruction, also contribute to the variance in a similar way, making estimation of the facets within the scene more challenging. 

\begin{figure}[ht]
    \centering
    {\phantomsubcaption\label{fig:robusta}}
    {\phantomsubcaption\label{fig:robustb}}
    \includegraphics[width=1\textwidth]{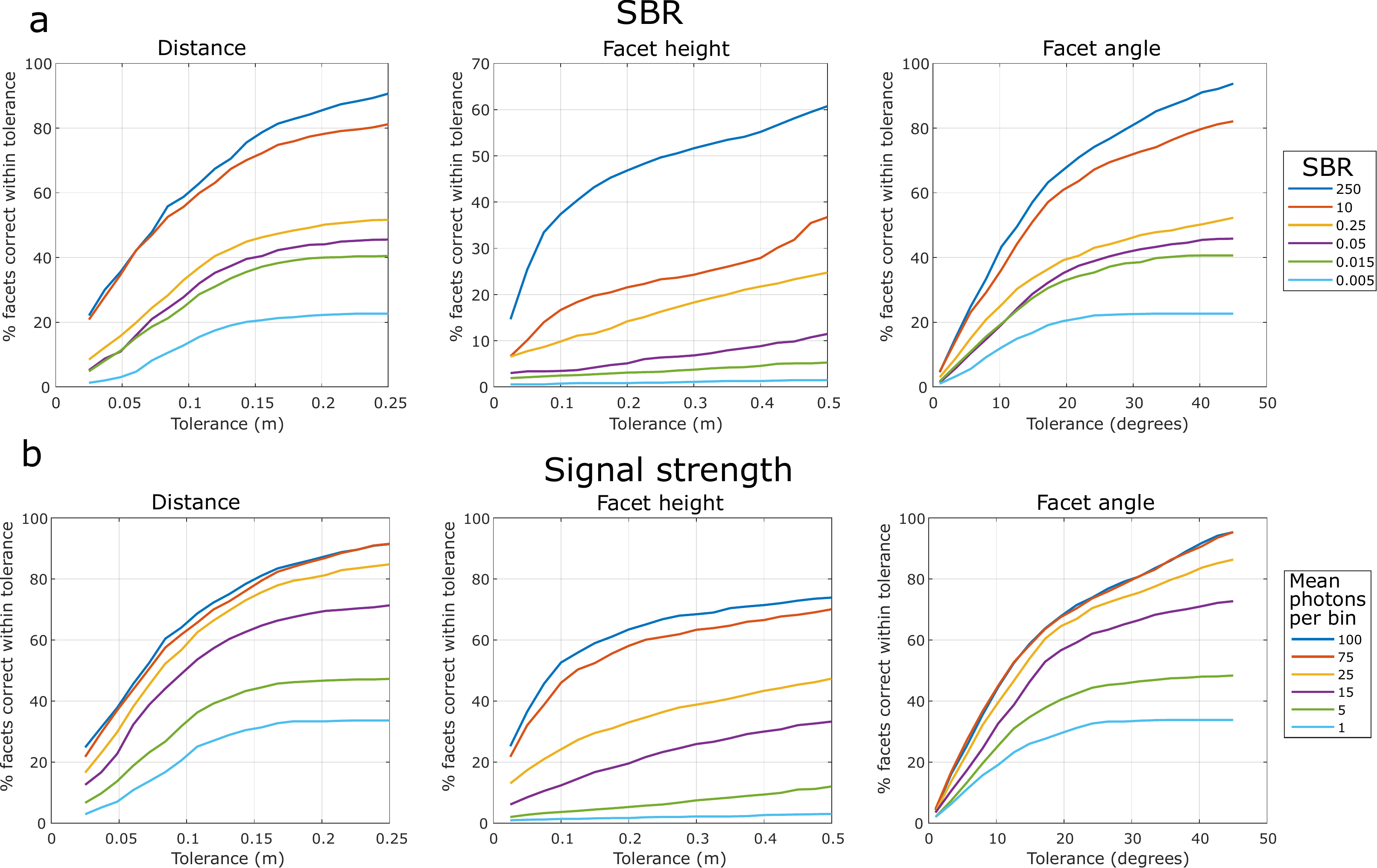}
    \caption{Simulated robustness results. \subref{fig:robusta} Parameter estimation success at varying signal-to-background ratios. 
    \subref{fig:robustb} Parameter estimation success with varying signal strength.}
    \label{fig:psbr}
\end{figure}

Supplementary Figure~\ref{fig:psbr}(a) shows simulated results for varying signal-to-background (SBR) ratios. SBR is defined as ratio of the total sum of signal photons from the hidden scene, to the total sum of background photons. A room with the same dimensions to that of the one in the main experimental results 
was simulated to generate ground truth measurement histograms. An appropriate, constant background level is added to each achieve the desired SBR (given an average photon count of 125 per time bin -- similar to the experimental measurements). The reconstruction algorithm was run using this measurement data (25 trials with new Poisson noise realizations), and the number of facets with parameters correctly identified within tolerances is presented.  

Supplementary Figure~\ref{fig:psbr}(b) shows similar results for varying signal strength. Signal strength here is defined by the average number of signal photon counts per time bin, given a time bin resolution of 16 ps and repetition rate of approximately 20 MHz. These were simulated with an SBR of 100.  

\begin{figure}[ht]
    \centering
    \includegraphics[trim={20mm 65mm 20mm 50mm},clip,width=0.5\textwidth]{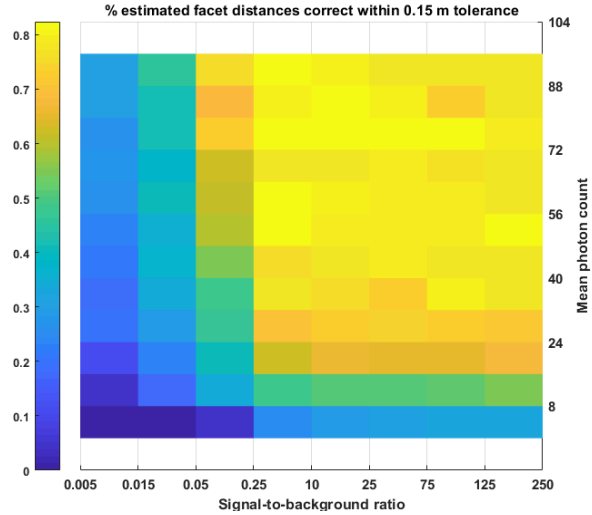}
    \caption{The number of facet distances estimated correctly within a tolerance of 0.15 m, as a function of SBR and signal strength.}
    \label{fig:sbr-sig}
\end{figure}

Supplementary Figure~\ref{fig:sbr-sig} shows the percentage of facet distances estimated correctly within a tolerance of 0.15 m, as a function of both SBR and signal strength. This can be useful to identify regimes in which the the system can operate with success.

\begin{figure}[ht]
\centering
\includegraphics[width=1\textwidth]{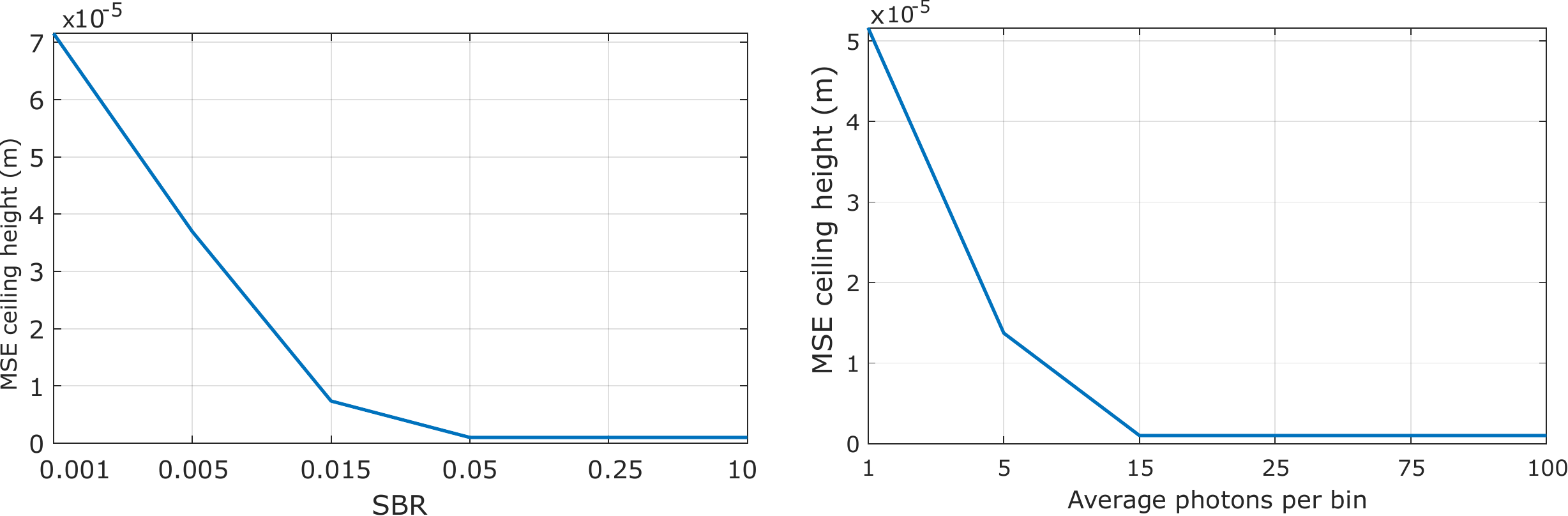}
\caption{Mean squared error for ceiling height estimation at varying SBR and signal strength.}
\label{fig:ceilingtest}
\end{figure}

Supplementary Figure~\ref{fig:ceilingtest} shows simulated results for the mean squared error of ceiling height estimation, using the same methods. We see that the mean squared error in general is extremely low, suggesting we can get accurate ceiling height estimations at a wide range of signal strengths and background strengths. This is due to the fact the ceiling is a strong signal, present in every measurement histogram. Every histogram is used in the ceiling height estimation at once, so given $N$ illumination positions, we have $N$ observations of the ceiling we use to form the estimate. 

